\documentclass[11pt,a4paper]{article}
\usepackage{bbm}
\pdfoutput=1
\usepackage{jheppub}
\usepackage{amsmath}
\usepackage{epsfig}
\usepackage{amssymb}
\usepackage{graphics}
\usepackage[active]{srcltx}
\usepackage{epstopdf}
\usepackage{subfigure}

\newcommand{\be}{\begin{equation}}
\newcommand{\ee}{\end{equation}}
\newcommand{\bea}{\begin{eqnarray}}
\newcommand{\eea}{\end{eqnarray}}
\newcommand{\bean}{\begin{eqnarray*}}
\newcommand{\eean}{\end{eqnarray*}}
\newcommand{\nn}{\nonumber \\}

\def\W #1{\widetilde{#1}}

\def\eref#1{(\ref{#1})}

\def\a{{\alpha}}

\def\b{{\beta}}

\def\Label#1{\label{#1}%
  \smash{\hbox to0pt{\raise1ex\hbox{\tiny[#1]}\hss}}}

\title{On differential operators and unifying relations for $1$-loop Feynman integrands}
\author[a]{Kang Zhou,}

\affiliation[a]{Center for Gravitation and Cosmology, College of Physical Science and Technology, Yangzhou University,\\
 No.180, Siwangting Road, Yangzhou, 225009, P.R. China.}

\date{\today}
\abstract{We generalize the unifying relations for tree amplitudes to the $1$-loop Feynman integrands.
By employing the $1$-loop CHY formula, we construct differential operators which transmute the $1$-loop gravitational Feynman integrand to Feynman integrands for a wide range
of theories, include Einstein-Yang-Mills theory, Einstein-Maxwell theory, pure Yang-Mills theory, Yang-Mills-scalar theory, Born-Infeld theory,
Dirac-Born-Infeld theory, bi-adjoint scalar theory, non-linear sigma model,
as well as special Galileon theory. The unified web at $1$-loop level is established. Under the well known unitarity cut, the $1$-loop level operators will factorize into two tree level operators. Such factorization is also discussed.
}


\begin{document}

\maketitle \flushbottom

\section{Introduction}
\label{secintro}

The past decades have revealed amazing relations and common structures within amplitudes of gauge and gravity theories,
such as the Kawai-Lewellen-Tye (KLT) relations \cite{KLT}, Bern-Carrasco-Johansson (BCJ) color-kinematics duality \cite{Bern:2008qj,Bern:2010ue,Bern:2010yg}, which
are invisible upon inspecting the traditional Feynman rules. These progresses hint the existence
of some long hidden unifying relations for on-shell amplitudes. The marvelous unity was first spelled out in
\cite{Cachazo:2014xea} by using the Cachazo-He-Yuan (CHY) formulations \cite{Cachazo:2013gna,Cachazo:2013hca,Cachazo:2013iea,Cachazo:2014nsa,Cachazo:2014xea}. In the CHY framework, different
theories are defined by different CHY integrands, while they found that CHY integrands for a wide range of theories can be generated from the CHY integrand for gravity theory\footnote{Here the
 gravity theory has to be understood in a generalized version,
i.e., Einstein gravity theory couples to a dilaton and two-forms.}, through the so called compactifying, squeezing,
as well as the generalized dimensional reduction procedures \cite{Cachazo:2014xea}.
More recently, similar unifying relations for
on-shell tree amplitudes of a variety of theories, based on constructing some Lorentz and gauge invariant differential operators, was proposed by Cheung, Shen and Wen \cite{Cheung:2017ems}.
By acting these differential operators, one can transmute the physical amplitude of a theory into the one of another theory.
The similarity between two unified webs implies the underlying connection between two approaches. This connection has been established in \cite{Zhou:2018wvn,Bollmann:2018edb,Zhou:2020umm},
by applying differential operators to CHY integrals for different theories.

It is natural to ask if these unifying relations can be generalized to the loop level. Motivated by the experience at the tree level,
we study this issue by considering the $1$-loop CHY formula. The $1$-loop CHY formula can be obtained via either the underlying ambitwistor string theory \cite{Adamo:2013tca,Mason:2013sva,Adamo:2013tsa,Casali:2014hfa,Geyer:2015bja,Geyer:2015jch,Adamo:2015hoa}, or the forward limit method \cite{He:2015yua,Cachazo:2015aol,Feng:2016nrf,Feng:2019xiq}.
In this paper, we focus on the latter one. Our basic idea can be summarized as follows. Suppose the tree amplitudes of theories $A$
and $B$ are connected by the operator ${\cal O}$ as ${\cal A}_B={\cal O}\,{\cal A}_A$, we seek the $1$-loop level operator
${\cal O}_\circ$ satisfying ${\cal O}_\circ\,{\cal F}\,{\cal A}_A={\cal F}\,{\cal O}\,{\cal A}_A$, where the operator ${\cal F}$ denotes taking the forward limit.
Since the $1$-loop Feynman integrands are obtained via
the forward limit as ${\bf I}_A=(1/\ell^2){\cal F}\,{\cal A}_A$ and ${\bf I}_B=(1/\ell^2){\cal F}\,{\cal A}_B$, one can conclude that the operator ${\cal O}_\circ$ transmutes the Feynman integrand in the desired manner ${\bf I}_B={\cal O}_\circ\,{\bf I}_A$.

The elegant structure of the tree and $1$-loop CHY formulae offers some advantages, which allow us to realize the above idea conveniently.
The tree amplitudes and the $1$-loop Feynman integrands in the CHY formulae are formulated as contour
integrals over auxiliary variables as
\bea
{\cal A}=\int d\mu\,{\cal I}^L{\cal I}^R\,,~~~~~~~~{\bf I}={1\over \ell^2}\int d\mu'\,{\cal F}\,\Big({\cal I}^L{\cal I}^R\Big)={1\over \ell^2}\int d\mu'\,\Big({\cal F}\,{\cal I}^L\Big)
\Big({\cal F}\,{\cal I}^R\Big)\,,~~~~\label{CHY-0}
\eea
respectively, where the auxiliary variables are localized by constraints from the so-called scattering equations. In these formulae, different theories are characterized by the so called CHY integrands ${\cal I}^L{\cal I}^R$ and ${\cal F}\,({\cal I}^L{\cal I}^R)$.
The tree level operators in \cite{Cheung:2017ems,Zhou:2018wvn,Bollmann:2018edb,Zhou:2020umm} are commutable with the CHY contour integrals $\int d\mu$, while the $1$-loop level operators ${\cal O}_\circ$ which will be constructed in this paper are commutable with the contour integral $\int d\mu'$.
Therefore, transmuting a Feynman integrand is equivalent to transmuting the associated CHY integrand. More explicitly, if two Feynman integrands ${\bf I}_A$ and ${\bf I}_B$ are related by an operator ${\cal O}_\circ$ as ${\bf I}_A={\cal O}_\circ\,{\bf I}_B$, analogous relation ${\cal F}\,({\cal I}^L{\cal I}^R)_B={\cal O}_\circ\,{\cal F}\,({\cal I}^L{\cal I}^R)_A$ for CHY integrands
must hold, and vice versa. Similarly, at the tree level, the connection ${\cal A}_A={\cal O}\,{\cal A}_B$ is equivalent to $({\cal I}^L
{\cal I}^R)_B={\cal O}\,({\cal I}^L
{\cal I}^R)_A$. Thus, one can study the unifying relations systematically
by acting operators on CHY integrands. Furthermore, the partial integrands ${\cal I}^L$ and ${\cal I}^R$ carry two independent sets of polarization vectors $\{\epsilon_i\}$ and $\{\W\epsilon_i\}$ respectively, and so do $({\cal F}\,{\cal I}^L)$ and
$({\cal F}\,{\cal I}^R)$. Thus one can define two classes of operators, where ${\cal O}^\epsilon_\circ$ and ${\cal O}^\epsilon$ depend on polarization vectors in $\{\epsilon_i\}$, while ${\cal O}^{\W\epsilon}_\circ$ and ${\cal O}^{\W\epsilon}$ depend on those in $\{\W\epsilon_i\}$. Then, for our purpose, it is sufficient to find the operator ${\cal O}^\epsilon_\circ$ which satisfies ${\cal O}^\epsilon_\circ\,{\cal F}\,{\cal I}^L={\cal F}\,{\cal O}^\epsilon\,{\cal I}^L$ (or equivalently the operator ${\cal O}^{\W\epsilon}_\circ$ satisfies ${\cal O}^{\W\epsilon}_\circ\,{\cal F}\,{\cal I}^R={\cal F}\,{\cal O}^{\W\epsilon}\,{\cal I}^R$). We will construct the $1$-loop differential operators ${\cal O}^{\epsilon}_\circ$ which satisfy the above requirement by considering the effects of the operators ${\cal O}^\epsilon$ at the tree level.

Using the method discussed above, we will find the $1$-loop differential operators which transmute the $1$-loop GR Feynman integrand to the Feynman integrands of a large verity of theories include Einstein-Yang-Mills (EYM) theory, Einstein-Maxwell (EM) theory, Born-Infeld (BI) theory,
Dirac-Born-Infeld (DBI) theory, special Galileon (SG) theory,  non-linear sigma
model (NLSM), as well as bi-adjoint scalar (BAS) theory, and establish the  unified web at $1$-loop level.

Under the well known unitarity cut, the $1$-loop Feynman integrand factorizes into two on-shell tree amplitudes.
Suppose the Feynman integrands of two theories are related as ${\bf I}_B={\cal O}_\circ\,{\bf I}_A$, under the unitarity cut,
the $1$-loop level operator ${\cal O}_\circ$ should also factorize into two tree level operators, due to the assumption ${\cal A}_B={\cal O}\,{\cal A}_A$ which serve as the foundation of our method. This property exhibits another connection between the tree and $1$-loop operators. We will also discuss such factorization of the $1$-loop level differential operators.

The remainder of this paper is organized as follows. In section.\ref{secreview}, we give a brief introduction to the tree level and $1$-loop level CHY formulae, the forward limit approach, as well as the tree level differential operators, which are crucial for subsequent discussions. In section.\ref{GR-YM-BAS}, we construct the $1$-loop level differential operator which transmutes the $1$-loop GR Feynman integrand to the YM Feynman integrand, and transmutes the YM Feynman integrand to the BAS Feynman integrand.
Then, in section.\ref{other theory} we apply the similar idea to other theories, and find operators which relate the GR Feynman integrand to Feynman integrand of a special case of the single trace EYM, the Feynman integrands of BI, NLSM, SG, EM and the extended EM that photons carry flavors, as well as DBI. The factorization of $1$-loop level differential operators under the unitarity cut will be studied in section.\ref{sectionfac-op}. Finally, we end with a summary and discussions in section.\ref{secconclu}, provide the $1$-loop level unified web of different theories.

\section{Background}
\label{secreview}

For reader's convenience, in this section we rapidly review the necessary background. In subsection.\ref{subsecCHY}, we give a brief introduction to the CHY formula at the tree level. Then, in subsection.\ref{forwardlimit}, we review the forward limit method, as well as the CHY formula at the $1$-loop level. Finally, the tree level differential operators, which link the tree level amplitudes of a wide range of theories together,
will be introduced in subsection.\ref{subsecOP}.

\subsection{Tree level CHY formula}
\label{subsecCHY}

In the CHY framework, tree amplitudes for $n$ massless particles in arbitrary dimensions arise from a multi-dimensional contour integral over
the moduli space of genus zero Riemann surfaces with $n$ punctures, ${\cal M}_{0,n}$ \cite{Cachazo:2013gna,Cachazo:2013hca,Cachazo:2013iea,Cachazo:2014nsa,Cachazo:2014xea}, formulated as
\bea
{\cal A}_n=\int d\mu_n\,{\cal I}^L(\{k,\epsilon,z\}){\cal I}^R(\{k,\W\epsilon,z\})\,,~~~~\label{CHY}
\eea
which possesses the M\"obius ${\rm SL}(2,\mathbb{C})$ invariance. Here $k_i$, $\epsilon_i$ and $z_i$ are the momentum, polarization vector, and puncture location for $i^{\rm th}$ external
particle, respectively. The measure part is defined as
\bea
d\mu_n\equiv{d^n z\over{\rm vol}\,{\rm SL}(2,\mathbb{C})}\prod_i{'}\delta({\cal E}_i)\,.
\eea
The $\delta$-functions impose the scattering equations
\bea
{\cal E}_i\equiv\sum_{j\in\{1,2,\ldots,n\}\setminus\{i\}}{k_i\cdot k_j\over z_{ij}}=0\,,
\eea
where $z_{ij}\equiv z_i-z_j$. The scattering equations define the map from the punctures on the moduli space ${\cal M}_{0,n}$ to vectors on the light cone, and fully localize the integral on
their solutions.
The measure part is universal ,while the integrand in \eref{CHY} depends on the theory under consideration. For any theory known to have a CHY representation, the corresponding integrand can be split into two
parts ${\cal I}^L$ and ${\cal I}^R$, as can be seen in \eref{CHY}. Either of them are weight-$2$ for each variable $z_i$
under the M\"obius transformation. In Table.\ref{tab:theories}, we list the tree level CHY integrands which will be used in this paper \cite{Cachazo:2014xea}\footnote{For theories contain gauge or flavor groups, we only show
the integrands for color-ordered partial amplitudes instead of full ones.}. Here EMf denotes the EM theory that photons carry flavors, sEYM and sYMS stand for the single trace EYM and YMS theories.
\begin{table}[!h]
    \begin{center}
        \begin{tabular}{c|c|c}
            Theory& ${\cal I}^L(k,\epsilon,z)$ & ${\cal I}^R(k,\W\epsilon,z)$ \\
            \hline
            GR & ${\bf Pf}'\Psi$ & ${\bf Pf}'{\Psi}$ \\
            sEYM & $PT(\overline{\sigma_1,\cdots,\sigma_m}){\bf Pf}[\Psi_l]_{n-m: n-m}$ & ${\bf Pf}'{\Psi}$ \\
            YM & $PT(\overline{\sigma_1,\cdots,\sigma_n})$ & ${\bf Pf}' \Psi$ \\
            EM & ${\bf Pf}'[\Psi]_{n-2m,2m:n-2m}{\bf Pf}[X]_{2m}$ & ${\bf Pf}'{\Psi}$ \\
            EMf & ${\bf Pf}'[\Psi]_{n-2m,2m;n-2m}{\bf Pf}[{\cal X}]_{2m}$
            & ${\bf Pf}'{\Psi}$ \\
            BI & $({\bf Pf}'A)^2$ & ${\bf Pf}' \Psi$ \\
            sYMS & $PT(\overline{\sigma_1,\cdots,\sigma_m}){\bf Pf}[\Psi_l]_{n-m: n-m}$ & $PT(\overline{\sigma'_1,\cdots,\sigma'_n})$ \\
            BAS & $PT(\overline{\sigma_1,\cdots,\sigma_n})$ & $PT(\overline{\sigma'_1,\cdots,\sigma'_n})$ \\
            NLSM & $({\bf Pf}' A)^2$ & $PT(\overline{\sigma'_1,\cdots,\sigma'_n})$ \\
            DBI  & ${\bf Pf}'[\Psi]_{n-2m,2m;n-2m}{\bf Pf}[{\cal X}]_{2m}$ & $({\bf Pf}' A)^2$ \\
            SG & $({\bf Pf}'A)^2$ & $({\bf Pf}' A)^2$ \\
        \end{tabular}
    \end{center}
    \caption{\label{tab:theories}Form of the integrands for various theories}
\end{table}

We now explain building blocks appearing in Table.\ref{tab:theories} in turn. There are five kinds of $n\times n$ matrixes
\bea
& &A_{ij} = \begin{cases} \displaystyle {k_{i}\cdot k_j\over z_{ij}} & i\neq j\,,\\
\displaystyle  ~~~ 0 & i=j\,,\end{cases} \qquad\qquad\qquad\qquad B_{ij} = \begin{cases} \displaystyle {\epsilon_i\cdot\epsilon_j\over z_{ij}} & i\neq j\,,\\
\displaystyle ~~~ 0 & i=j\,,\end{cases} \nn
& &C_{ij} = \begin{cases} \displaystyle {k_i \cdot \epsilon_j\over z_{ij}} &\quad i\neq j\,,\\
\displaystyle -\sum_{l=1,\,l\neq j}^n\hspace{-.5em}{k_l \cdot \epsilon_j\over z_{lj}} &\quad i=j\,,\end{cases}
\label{ABCmatrix}
\eea
and
\bea
X_{ij}=\begin{cases} \displaystyle \frac{1}{z_{ij}} & i\neq j\,,\\
\displaystyle ~ ~ 0 & i=j\,,\end{cases} \qquad\qquad\qquad\qquad
{\cal X}_{ij}=\begin{cases} \displaystyle \frac{\delta^{I_i,I_j}}{z_{ij}} & i\neq j\,,\\
\displaystyle ~ ~ 0 & i=j\,.\end{cases}
\eea
where $\delta^{I_i,I_j}$ forbids the interaction between particles with different flavors. To clarify the dimension, we denote the $n\times n$ matrixes $X$ and ${\cal X}$ as $[X]_n$, $[{\cal X}]_n$. The $2n\times2n$ antisymmetric matrix $\Psi$ is given by
\bea\label{Psi}
\Psi = \left(
         \begin{array}{c|c}
           ~~A~~ &  ~~C~~ \\
           \hline
           -C^{\rm T} & B \\
         \end{array}
       \right)\,.
\eea
The definition of $\Psi$ can be generalized to the $(2a+b)\times(2a+b)$
case $[\Psi]_{a,b:a}$  as
\bea
[\Psi]_{a,b:a}=\left(
         \begin{array}{c|c}
           ~~A_{(a+b)\times (a+b)}~~ &  C_{(a+b)\times a} \\
           \hline
            -C^{\rm T}_{a\times (a+b)} & B_{a\times a} \\
         \end{array}
       \right)\,,~~~~\label{psi-aba}
\eea
here $A$ is a $(a+b)\times (a+b)$ matrix, $C$ is a $(a+b)\times a$ matrix, and $B$ is a $a\times a$ matrix.
The definitions of elements of $A$, $B$ and $C$ are the same as before.

The notation ${\bf Pf}$ stands for the polynomial called Pfaffian. For a $2n\times 2n$ skew symmetric matrix $S$, Pfaffian is defined as
\bea
{\bf Pf}S={1\over 2^n n!}\sum_{\sigma\in S_{2n}} {\bf sgn}(\sigma)\prod_{i=1}^n\,a_{\sigma(2i-1),\sigma(2i)}\,,~~~\label{pfa-1}
\eea
where $S_{2n}$ is the permutation group of $2n$ elements and ${\bf sgn}(\sigma)$ is the signature of $\sigma$.
More explicitly, let $\Pi$ be the set of all partitions of $\{1,2,\cdots, 2n\}$ into pairs without regard to the order.
An element $\a$ in $\Pi$ can be written as
\bea
\a=\{(i_1,j_1),(i_2,j_2),\cdots,(i_n,j_n) \}\,,
\eea
with $i_k<j_k$ and $i_1<i_2<\cdots<i_n$. Now let
\bea
\sigma_{\a} = \left(
         \begin{array}{c}
           ~~~1~~~ 2~~~3~~~4~~\cdots~2n-1~~2n~~ \\
           \,\,i_1~~j_1~~i_2~~j_2~~\cdots~~~i_n~~~~~~j_n \\
         \end{array}
       \right)
\eea
be the associated permutation of the partition $\a$. If we define
\bea
S_{\a}={\bf sgn}(\sigma_{\a})\,a_{i_1j_1}a_{i_2j_2}\cdots a_{i_nj_n}\,,
\eea
then the Pfaffian of the matrix $S$ is given as
\bea
{\bf Pf}S=\sum_{\a\in\Pi}S_{\a}\,.~~~~~\label{pfa}
\eea
From the \eref{pfa} one can observe that in every term $S_{\a}$
of the Pfaffian, each number of $\{1,2,\cdots,2n\}$, as the subscript of the matrix element, will appear once and only once. This observation is simple but useful for latter discussions.

With the definition of Pfaffian provided above, the reduced Pfaffian of the matrix $\Psi$ is defined as ${\bf Pf}'\Psi={(-)^{i+j}\over z_{ij}}{\bf Pf}\Psi^{i,j}_{i,j}$,
where the notation $\Psi^{i,j}_{i,j}$ means the rows and columns $i$, $j$ of the matrix $\Psi$
have been deleted (with $1\leq i,j\leq n$). It can be proved that this definition is independent of the choice of $i$ and $j$.
Analogous notation holds for ${\bf Pf}'A$.
The reduced Pfaffian ${\bf Pf}'[\Psi]_{a,b:a}$
is defined in the same manner. With the definition of the reduced Pfaffian, one can observe that each polarization vector $\epsilon_i$
appears once and only once in each term of the reduced Pfaffian.

Finally, the Parke-Taylor factor for ordering $\sigma$ is given as
\bea
PT(\overline{\sigma_1,\cdots,\sigma_n)}={1\over z_{\sigma_1\sigma_2}z_{\sigma_2\sigma_3}\cdots z_{\sigma_{n-1}\sigma_n}z_{\sigma_n\sigma_1}}\,,
\eea
it implies the color ordering $\overline{\sigma_1,\cdots,\sigma_n}$ for the partial amplitude. Throughout this paper, we use $\overline{\sigma_1,\cdots,\sigma_m}$ to denote the color ordering among $m$ elements $\sigma_i$.

\subsection{Forward limit method and $1$-loop CHY formula}
\label{forwardlimit}

The $1$-loop CHY formula can be obtained via the so called forward limit method. The forward limit is reached as follows:
\begin{itemize}
\item Consider a
$(n+2)$-point tree amplitude ${\cal A}(k_1,\cdots,k_n,k_+,k_-)$ including $n$ massless legs with momenta in $\{k_1,\cdots,k_n\}$ and two massive legs with $k_+^2=k_-^2=m^2$;
\item Take the limit $k_{\pm}\to \pm \ell$, and glue the two corresponding legs together;
\item Sum over all allowed internal states of the internal particle with loop momenta $\ell$, such as polarization
vectors or tensors, colors, flavors, and so on\footnote{For theories include gauge or flavor groups, we only discuss the color ordered partial amplitudes in this paper, thus the summations over colors or flavors are hidden. We will encounter the summation over flavors when considering the EMf theory that photons carry flavors.}.
\end{itemize}
Roughly speaking, the obtained object, times the factor $1/\ell^2$ as
\bea
{1\over \ell^2}\,{\cal F}\,{\cal A}(k_1,\cdots,k_n,k_+^{h_+},k_-^{h_-})={1\over \ell^2}\,\sum_{h}\,{\cal A}(k_1,\cdots,k_n,\ell^{h},-\ell^{\bar{h}})\,,
\eea
contributes to the $1$-loop Feynman integrand ${\bf I}(k_1,\cdots,k_n)$. Here we introduced the forward limit operator ${\cal F}$ to denote the operation of taking forward limit. In this paper, we denote the $1$-loop Feynman integrands by ${\bf I}$, to
distinguish them from the CHY integrands ${\cal I}$.

Now we introduce the relation among ${\cal F}\,{\cal A}(k_1,\cdots,k_n,k_+^{h_+},k_-^{h_-})$ and ${\bf I}(k_1,\cdots,k_n)$, in the CHY framework. For the tree amplitude ${\cal A}(k_1,\cdots,k_n,k_+,k_-)$ with two massive legs $k_+^2=k_-^2=m^2$, there are $(n+2)$ scattering equations, given as
\bea
& &{\cal E}_i\equiv\sum_{j\in\{1,2,\ldots,n\}\setminus\{i\}}{k_i\cdot k_j\over z_{ij}}+{k_i\cdot k_+\over z_{i+}}+{k_i\cdot k_-\over z_{i-}}=0\,,~~~~~~~~i\in\{1,\cdots,n\}\nn
& &{\cal E}_+\equiv\sum_{j=1}^n{k_+\cdot k_j\over z_{+j}}+{k_+\cdot k_-+m^2\over z_{+-}}=0\,,~~~~~~~~
{\cal E}_-\equiv\sum_{j=1}^n{k_-\cdot k_j\over z_{-j}}+{k_+\cdot k_-+m^2\over z_{-+}}=0\,.~~~~\label{SE-tree-2massive}
\eea
In the limit $k_\pm\to \pm \ell$, these equations behave as
\bea
& &{\cal E}_i\equiv\sum_{j\in\{1,2,\ldots,n\}\setminus\{i\}}{k_i\cdot k_j\over z_{ij}}+{k_i\cdot \ell\over z_{i+}}-{k_i\cdot \ell\over z_{i-}}=0=0\,,~~~~~~~~i\in\{1,\cdots,n\}\nn
& &{\cal E}_+\equiv\sum_{j=1}^n{\ell\cdot k_j\over z_{+j}}=0\,,~~~~~~~~
{\cal E}_-\equiv\sum_{j=1}^n{-\ell\cdot k_j\over z_{-j}}=0\,.~~~~\label{SE-loop}
\eea
These $1$-loop scattering equations yield the massive propagators $1/((\ell+K)^2-\ell^2)$ in the loop, rather than
the desired massless ones $1/(\ell+K)^2$. However, these massive propagators relate to the massless ones through the
well known partial fraction identity
\bea
{1\over D_1\cdots D_m}=\sum_{i=1}^m\,{1\over D_i}\Big[\prod_{j\neq i}\,{1\over D_j-D_i}\Big]\,,
\eea
which implies
\bea
{1\over \ell^2(\ell+K_1)^2(\ell+K_1+K_2)^2\cdots (\ell+K_1+\cdots+K_{m-1})^2}\simeq{1\over \ell^2}\sum_{i=1}^m\Big[\prod_{j= i}^{i+m-2}\,{1\over (\ell+K_i+\cdots+K_j)^2-\ell^2}\Big]\,.~~~~\label{partial-fraction}
\eea
For each individual term at the r.h.s of the above relation, we have shifted the loop momentum without alternating the result of Feynman integral.
Here $\simeq$ means the l.h.s and r.h.s are not equivalent to each other at the integrand level, but are equivalent at the integration level.
We emphasize that the l.h.s of \eref{partial-fraction} is the standard propagators in the loop for an individual diagram, contributes to ${\bf I}(k_1,\cdots,k_n)$, while each term at the r.h.s can be obtained via the forward limit method.

Thus, to construct the correct $1$-loop Feynman integrand ${\bf I}(k_1,\cdots,k_n)$ via the $1$-loop scattering equations in \eref{SE-loop}, one need to cut each propagator in the loop once, and sum over all resulting objects, as required by the partial fraction
relation \eref{partial-fraction}. For the amplitude without any color ordering, this requirement is satisfied automatically when summing over
all possible Feynman diagrams, thus we have
\bea
{\bf I}(k_1,\cdots,k_n)={1\over \ell^2}\,{\cal F}\,{\cal A}(k_1,\cdots,k_n,k_+^{h_+},k_-^{h_-})\,.~~~~\label{I-unclolor}
\eea
For the color ordered amplitude, this requirement is satisfied by summing over color orderings cyclically, namely
\bea
{\bf I}(\overline{\sigma_1,\cdots,\sigma_n})={1\over \ell^2}\,\sum_{i\in\{1,\cdots,n\}}\,{\cal F}\,{\cal A}(\overline{+,\sigma_i,\cdots,\sigma_{i-1},-})\,.~~~~\label{I-color}
\eea
The integrands ${\bf I}(k_1,\cdots,k_n)$ and ${\bf I}(\overline{\sigma_1,\cdots,\sigma_n})$ provided in \eref{I-unclolor}
and \eref{I-color} are not the original $1$-loop Feynman integrands written via Feynman rules, since the loop momenta have been
shifted in each term. But we still regard them as Feynman integrands, since they are equivalent to the original ones at the integration level.
In other words, the $1$-loop Feynman integrand should be understood as a class of integrands which give the same result after doing the integration.

As an equivalent interpretation, the forward limit method can also be understood from the dimensional reduction point of view, as studied in \cite{Cachazo:2015aol}.

Let us take a brief look at the CHY integrand at the $1$-loop level.
In the CHY framework, the forward limit operator ${\cal F}$ acts on the $(n+2)$-point tree amplitude as follows
\bea
{\cal F}\,{\cal A}_{n+2}&=&{\cal F}\,\int d\mu_{n+2}\,{\cal I}^L(\{k,\epsilon,z\}){\cal I}^R(\{k,\W\epsilon,z\})\nn
&=&\int d\mu'_{n+2}\,\Big({\cal F}\,{\cal I}^L(\{k,\epsilon,z\})\Big)\Big({\cal F}\,{\cal I}^R(\{k,\W\epsilon,z\})\Big)\,,~~~~\label{1loop-CHY}
\eea
where the measure $d\mu'_{n+2}$ is generated from $d\mu_{n+2}$ by turning the scattering equations in \eref{SE-tree-2massive} to those in \eref{SE-loop}.
Thus the $1$-loop CHY integrand is determined by
\bea
{\cal I}^{L}_\circ(\{k,\epsilon,z\})={\cal F}\,{\cal I}^L(\{k,\epsilon,z\})\,,~~~~~~~~
{\cal I}^{R}_\circ(\{k,\epsilon,z\})={\cal F}\,{\cal I}^R(\{k,\epsilon,z\})\,.
\eea
Using this statement, the $1$-loop CHY integrands for GR, YM and BAS are given in Table \ref{tab:1-loop-theories}.
\begin{table}[!h]
    \begin{center}
        \begin{tabular}{c|c|c}
            Theory& ${\cal I}^L_\circ(k,\epsilon,z)$ & ${\cal I}^R_\circ(k,\W\epsilon,z)$ \\
            \hline
            GR & ${\cal F}\,{\bf Pf}'\Psi$ & ${\cal F}\,{\bf Pf}'{\Psi}$ \\
            YM & $PT_\circ(\overline{\sigma_1,\cdots,\sigma_n})$ & ${\cal F}\,{\bf Pf}' \Psi$ \\
            BAS & $PT_\circ(\overline{\sigma_1,\cdots,\sigma_n})$ & $PT_\circ(\overline{\sigma'_1,\cdots,\sigma'_n})$ \\
        \end{tabular}
    \end{center}
    \caption{\label{tab:1-loop-theories}$1$-loop CHY integrands }
\end{table}
The ingredient ${\cal F}\,{\bf Pf}' \Psi$ is obtained by
\bea
{\cal F}\,{\bf Pf}'{\Psi}=\sum_r\,{\bf Pf}'{\Psi}^r_\ell\,,
\eea
due to the definition of the operator ${\cal F}$.
Here $\Psi$ is a $2(n+2)\times2(n+2)$ matrix constructed by $\{k_1,\cdots,k_n,k_+,k_-\}$ and $\{\epsilon_1,\cdots,\epsilon_n,\epsilon_+,\epsilon_-\}$, while ${\Psi}^r_\ell$ is obtained from $\Psi$ by setting $k_+=-k_-=\ell$
and $\epsilon_+=\epsilon^r_+$, $\epsilon_-=\epsilon^r_-=(\epsilon^r_+)^\dag$. The summation is over all allowed $\epsilon^r_+$.
For simplicity, we assume the reduced Pfaffian is evaluated by removing rows and columns correspond to $k_+$ and $k_-$.
From now on, when referring to ${\bf Pf}'{\Psi}$, ${\bf Pf}'{\Psi}^r_\ell$ and ${\cal F}\,{\bf Pf}' \Psi$, we always mean the objects introduced above.
The $1$-loop Parke-Taylor factor $PT_\circ(\overline{\sigma_1,\cdots,\sigma_n})$ is obtained by summing over tree Parke-Taylor
factors cyclically,
\bea
PT_\circ(\overline{\sigma_1,\cdots,\sigma_n})=\sum_{i\in\{1,\cdots,n\}}\,PT(\overline{+,\sigma_i,\cdots,\sigma_{i-1},-})\,,~~~~\label{PT-C}
\eea
as required by the rule in \eref{I-color}. Notice that since the Parke-Taylor factor is defined only through the coordinates of punctures, we have
\bea
{\cal F}\,PT(\overline{+,\sigma_i,\cdots,\sigma_{i-1},-})=PT(\overline{+,\sigma_i,\cdots,\sigma_{i-1},-})\,.
\eea
The tree Parke Taylor factor $PT(\cdots)$ in \eref{PT-C} should be understood as
${\cal F}\,PT(\cdots)$.
The integrands in Table \ref{tab:1-loop-theories} can be found in \cite{Geyer:2015jch,Geyer:2017ela,He:2015yua,Cachazo:2015aol}.

The $1$-loop CHY formula in \eref{1loop-CHY} suffer from the divergence in the forward limit. It was observed in \cite{He:2015yua}
that the solutions of $1$-loop scattering equations separate into three sectors which are called regular, singular ${\bf I}$ and singular ${\bf II}$,
according to the behavior of punctures $z_{\pm}$ in the limit $k_++k_-\to 0$. In this paper, we will bypass this subtle and crucial point by employing the conclusion in \cite{Cachazo:2015aol}, which can be summarized as follows: as long as the CHY integrand is homogeneous in $\ell^\mu$,
the singular solutions contribute to the scaleless integrals which vanish under the dimensional regularization. The homogeneity is manifest for the Parke-Taylor factor. On the other hand, for ${\cal F}\,{\bf Pf}'{\Psi}$, the only place that can violate the homogeneity in $\ell^\mu$ is the diagonal elements in the matrix $C$, since the deleted rows and columns are chosen to be $k_+$ and $k_-$. For singular solutions we have $z_+=z_-$, then it is direct to observe that the dependence on $\ell^\mu$ exactly cancel away, left with a homogeneous CHY integrand. The CHY integrands for BI, NLSM, SG, EM and DBI,  which will be considered in latter sections, are also homogeneous in $\ell^\mu$, as can be verified straightforwardly.
This observation allows us to ignore the problem of singular solutions.

\subsection{Differential operators at tree level}
\label{subsecOP}

The differential operators introduced by Cheung, Shen and Wen transmute tree amplitudes of one theory to those of other theories \cite{Cheung:2017ems,Zhou:2018wvn,Bollmann:2018edb}. Three kinds of basic operators are defined as follows:
\begin{itemize}
\item (1) Trace operator:
\bea
{\cal T}^\epsilon[\overline{i,j}]\equiv \partial_{\epsilon_i\cdot\epsilon_j}\,,
\eea
where $\epsilon_i$ is the polarization vector of $i^{\rm th}$ external leg. The up index $\epsilon$ means the operators are defined through polarization vectors in $\{\epsilon_i\}$.
\item (2) Insertion operator:
\bea
{\cal I}^\epsilon_{ikj}\equiv \partial_{\epsilon_k\cdot k_i}-\partial_{\epsilon_k\cdot k_j}\,,~~~~\label{defin-insertion}
\eea
where $k_i$ denotes the momentum of the $i^{\rm th}$ external leg. When applying to physical amplitudes, the insertion operator ${\cal I}^\epsilon_{ik(i+1)}$ inserts the external leg $k$ between external legs $i$ and $(i+1)$ in the color-ordering $(\cdots,i,i+1,\cdots)$. For general ${\cal I}^\epsilon_{ikj}$ with $i<j$, one can use the definition \eref{defin-insertion} to decompose ${\cal I}^\epsilon_{ikj}$ as
\bea
{\cal I}^\epsilon_{ikj}={\cal I}^\epsilon_{ik(i+1)}+{\cal I}^\epsilon_{(i+1)k(i+2)}+\cdots+{\cal I}^\epsilon_{(j-1)kj}\,.
\eea
In the above expression, each ${\cal I}^\epsilon_{ak(a+1)}$ on the RHS can be interpreted as inserting the leg $k$ between $a$ and $(a+1)$.
Consequently, the effect of ${\cal I}^\epsilon_{ikj}$ can be understood as inserting $k$ between $i$ and $j$ in the color-ordering $(\cdots,i,\cdots,j,\cdots)$, and summing over all possible positions together.
\item (3) Longitudinal operator:
\bea
{\cal L}^\epsilon_i\equiv \sum_{j\neq i}\,(k_i\cdot k_j)\partial_{\epsilon_i\cdot k_j}\,,~~~~~~~~
{\cal L}^\epsilon_{ij}\equiv -(k_i\cdot k_j)\partial_{\epsilon_i\cdot \epsilon_j}\,.
\eea
\end{itemize}

By using products of these three kinds of basic operators, one can transmute amplitudes of one theory into those of other theories. Three combinatory operators which are products of basic operators are defined as follows:
\begin{itemize}
\item (1) For a length-$m$ ordered set $\overline{\pmb{\sigma}}_m=\{\sigma_1,\cdots,\sigma_m\}$ of external particles, the operator ${\cal T}^\epsilon[\overline{\pmb{\sigma}}_m]$
is given as\footnote{In this paper, we adopt the convention that the operator at l.h.s acts after the operator at r.h.s. From the mathematical
point of view, the order of operators is irrelevant, since all operators are commutative with each other. We choose the order of operators in the
definition to emphasize the interpretation of each one.}
\bea
{\cal T}^\epsilon[\overline{\pmb{\sigma}}_m]\equiv \Big(\prod_{i=2}^{m-1}\,{\cal I}^\epsilon_{\sigma_1\sigma_i\sigma_{i+1}}\Big)\,{\cal T}^\epsilon[\overline{\sigma_1,\sigma_m}]\,.~~~~\label{defin-T}
\eea
It fixes $\sigma_1$ and $\sigma_m$ at two ends of the color-ordering via the operator ${\cal T}^\epsilon[\overline{\sigma_1,\sigma_m}]$, and inserts other elements between them by insertion operators.
The operator ${\cal T}^\epsilon[\overline{\pmb{\sigma}}_m]$ is also called the trace operator since it generates the color-ordering $\overline{\sigma_1,\sigma_2,\cdots,\sigma_m}$. The interpretation of insertions operators indicates that ${\cal T}^\epsilon[\overline{\pmb{\sigma}}_m]$ has various equivalent formulae, for example
\bea
& &{\cal T}^\epsilon[\overline{\pmb{\sigma}}_m]= \Big(\prod_{i=m-1}^{2}\,{\cal I}^\epsilon_{\sigma_{i-1}\sigma_i\sigma_m}\Big)\,{\cal T}^\epsilon[\overline{\sigma_1,\sigma_m}]\,,\nn
& &{\cal T}^\epsilon[\overline{\pmb{\sigma}}_m]= \Big(\prod_{i=3}^{m-3}\,{\cal I}^\epsilon_{\alpha_2\alpha_i\alpha_{i+1}}\Big)\, {\cal I}^\epsilon_{\sigma_{2}\sigma_{m-2}\sigma_{m-1}}\,{\cal I}^\epsilon_{\sigma_{1}\sigma_2\sigma_{m-1}}\,{\cal I}^\epsilon_{\sigma_{m-1}\sigma_m\sigma_1}\,{\cal T}^\epsilon[\overline{\sigma_1,\sigma_{m-1}}]\,,
\eea
and so on. The second example provided above shows that it is not necessary to choose the first operator to be ${\cal T}^\epsilon[\overline{\sigma_1,\sigma_m}]$. In other words, two reference legs in the color ordering can be chosen arbitrary.
\item (2) For $n$-point amplitudes, the operator ${\cal L}^\epsilon$
is defined as
\bea
{\cal L}^\epsilon\equiv\prod_i\,{\cal L}^\epsilon_i,~~~~~~~~\bar{{\cal L}}^\epsilon\equiv\sum_{\rho\in{\rm pair}}\,\prod_{i,j\in\rho}\,{\cal L}^\epsilon_{ij}\,.~~~~\label{defin-L}
\eea
Two definitions ${\cal L}^\epsilon$ and $\bar{{\cal L}}^\epsilon$ are not equivalent to each other at the algebraic level. However, when acting on proper on-shell
physical amplitudes, two combinations
${\cal L}^\epsilon\cdot{\cal T}^\epsilon[\overline{a,b}]$ and $\bar{{\cal L}}^\epsilon\cdot{\cal T}^\epsilon[\overline{a,b}]$, with subscripts of ${\cal L}^\epsilon_i$ and ${\cal L}^\epsilon_{ij}$
run through all nodes in $\{1,2,\cdots,n\}\setminus\{a,b\}$, give the same effect which can be interpreted physically.
\item (3) For a length-$2m$ set $\pmb{I}$, the operator ${\cal T}^\epsilon_{{\cal X}_{2m}}$ is defined as
\bea
{\cal T}^\epsilon_{{\cal X}_{2m}}\equiv\sum_{\rho\in{\rm pair}}\,\prod_{i_k,j_k\in\rho}\,\delta_{I_{i_k}I_{j_k}}{\cal T}^\epsilon[\overline{i_k,j_k}]\,,~~~~\label{TX1}
\eea
where $\delta_{I_{i_k}I_{j_k}}$ forbids the interaction between particles with different flavors. For the special case $2m$ particles do not carry any flavor, the operator ${\cal T}^\epsilon_{X_{2m}}$ is defined by removing $\delta_{I_{i_k}I_{j_k}}$,
\bea
{\cal T}^\epsilon_{X_{2m}}\equiv\sum_{\rho\in{\rm pair}}\,\prod_{i_k,j_k\in\rho}\,{\cal T}^\epsilon[\overline{i_k,j_k}]\,.~~~\label{TX2}
\eea
\end{itemize}
The explanation for the notation $\sum_{\rho\in{\rm pair}}\,\prod_{i_k,j_k\in\rho}$ is in order. Let $\Gamma$ be the set of all partitions of the set $\{1,2,\cdots, 2m\}$ into pairs without regard to the order.
An element in $\Gamma$ can be written as
\bea
\rho=\{(i_1,j_1),(i_2,j_2),\cdots,(i_m,j_m)\}\,,
\eea
with conditions $i_i<i_2<\cdots<i_m$ and $i_t<j_t,\,\forall t$. Then, $\prod_{i_k,j_k\in\rho}$ stands for the product of ${\cal T}^\epsilon[\overline{i_k,j_k}]$
for all pairs $(i_k,j_k)$ in $\rho$, and $\sum_{\rho\in{\rm pair}}$ denotes the summation over all partitions.

The combinatory operators exhibited above unify tree amplitudes of a wide range of theories together, by translating the GR amplitudes into amplitudes of other theories, formally expressed as
\bea
{\cal A}={\cal O}^\epsilon{\cal O}^{\W\epsilon}{\cal A}^{\epsilon,\W\epsilon}_{\rm GR}\,.~~~~\label{fund-uni-diff}
\eea
Operators ${\cal O}^\epsilon$ and ${\cal O}^{\W\epsilon}$ for different theories, which will be used in this paper, are listed in Table \ref{tab:unifying}.
\begin{table}[!h]
\begin{center}
\begin{tabular}{c|c|c}
Amplitude& ${\cal O}^\epsilon$  & ${\cal O}^{\W\epsilon}$ \\
\hline
${\cal A}_{{\rm GR}}^{\epsilon,\W\epsilon}(\pmb{H}_n)$ & $\mathbb{I}$ & $\mathbb{I}$  \\
${\cal A}_{{\rm sEYM}}^{\epsilon,\W\epsilon}(\overline{\pmb{\sigma}}_m;\pmb{H}_{n-m})$ & $\mathbb{I}$ & ${\cal T}^{\W\epsilon}[\overline{\pmb{\sigma}}_m]$  \\
${\cal A}_{{\rm EMf}}^{\epsilon,\W\epsilon}(\pmb{P}_{2m};\pmb{H}_{n-2m})$ & $\mathbb{I}$ & ${\cal T}^{\W\epsilon}_{{\cal X}_{2m}}$  \\
${\cal A}_{{\rm EM}}^{\epsilon,\W\epsilon}(\pmb{P}_{2m};\pmb{H}_{n-2m})$ & $\mathbb{I}$ & ${\cal T}^{\W\epsilon}_{X_{2m}}$  \\
${\cal A}_{{\rm BI}}^\epsilon(\pmb{P}_n)$ & $\mathbb{I}$ & ${\cal L}^{\W\epsilon}\,{\cal T}^{\W\epsilon}[\overline{a,b}]$ \\
${\cal A}_{{\rm YM}}^\epsilon(\overline{\pmb{\sigma}}_n)$ & $\mathbb{I}$ & ${\cal T}^{\W\epsilon}[\overline{\pmb{\sigma}}_n]$  \\
${\cal A}_{{\rm sYMS}}^{\W\epsilon}(\overline{\pmb{\sigma}}_m;\pmb{G}_{n-m}|\overline{\pmb{\sigma}'}_n)$ & ${\cal T}^{\epsilon}[\overline{\pmb{\sigma}'}_n]$ & ${\cal T}^{\W\epsilon}[\overline{\pmb{\sigma}}_m]$ \\
${\cal A}_{{\rm NLSM}}(\overline{\pmb{\sigma}'}_n)$ & ${\cal T}^{\epsilon}[\overline{\pmb{\sigma}'}_n]$ & ${\cal L}^{\W\epsilon}\,{\cal T}^{\W\epsilon}[\overline{a,b}]$ \\
${\cal A}_{{\rm BAS}}(\overline{\pmb{\sigma}}_n|\overline{\pmb{\sigma}'}_n)$ &  ${\cal T}^{\epsilon}[\overline{\pmb{\sigma}'}_n]$ & ${\cal T}^{\W\epsilon}[\overline{\pmb{\sigma}}_n]$ \\
${\cal A}_{{\rm DBI}}^{\W\epsilon}(\pmb{S}_{2m};\pmb{P}_{n-2m})$ &${\cal L}^{\epsilon}\,{\cal T}^{\epsilon}[\overline{a',b'}]$ & ${\cal T}^{\W\epsilon}_{{\cal X}_{2m}}$ \\
${\cal A}_{{\rm SG}}(\pmb{S}_n)$ &  ${\cal L}^{\epsilon}\,{\cal T}^{\epsilon}[\overline{a',b'}]$ & ${\cal L}^{\W\epsilon}\,{\cal T}^{\W\epsilon}[\overline{a,b}]$ \\
\end{tabular}
\end{center}
\caption{\label{tab:unifying}Unifying relations for differential operators at tree level.}
\end{table}
In this table, all amplitudes include $n$ external particles. The symbol $\mathbb{I}$ stands for the identical operator. Notations $\pmb{H}_a$, $\pmb{P}_a$, $\pmb{G}_a$ and $\pmb{S}_a$
denote sets of gravitons, photons, gluons and scalars respectively, where the subscript denotes the length of the set. A bold number or letter stands for a set, and $\pmb{\bar{}}$ denotes that the set is ordered. We use ${\cal A}_{{\rm sYMS}}^{\W\epsilon}(\overline{\pmb{\sigma}}_m;\pmb{G}_{n-m}|\overline{\pmb{\sigma}'}_n)$ as the example to explain notations $|$, and $;$.
The additional color ordering among all external particles is presented at the r.h.s of the notation $|$, such as $\overline{\pmb{\sigma}'}_n$ among all scalars and gluons in the example. Notation $;$ is used to separate different sets of external particles, particles on the l.h.s of $;$ carry lower spin. In our example, the l.h.s of $;$ is the set of scalars while the r.h.s is the set of gluons. The up index of
${\cal A}$ denotes the polarization vectors of external particles. In the cases amplitudes include external gravitons, the rule is: the previous polarization vectors are carried by all particles,
while the later ones are only carried by gravitons. For instance, in the notation ${\cal A}_{{\rm EMf}}^{\epsilon,\W\epsilon}(\pmb{P}_{2m};\pmb{H}_{n-2m})$, $\epsilon_i$ are carried by both photons and gravitons, while $\W\epsilon_i$ are only carried by gravitons.

In Table \ref{tab:unifying}, two sectors of operators labeled by polarization vectors $\epsilon$ and $\W\epsilon$ are exchangeable. As an example, YM amplitudes carry the polarization vectors $\W\epsilon$ can be generated by
\bea
{\cal A}^{\W\epsilon}_{\rm YM}(\overline{\pmb{\sigma}}_n)={\cal T}^\epsilon[\overline{\pmb{\sigma}}_n]{\cal A}^{\epsilon,\W\epsilon}_{\rm GR}(\pmb{H}_n)\,.
\eea
All relations between amplitudes of different theories can be extracted from Table \ref{tab:unifying}. For example, from relations
\bea
& &{\cal A}^\epsilon_{\rm BI}(\pmb{P}_n)={\cal L}^{\W\epsilon}\,{\cal T}^{\W\epsilon}[\overline{a,b}]{\cal A}^{\epsilon,\W\epsilon}_{\rm GR}(\pmb{H}_n)\,,\nn
& &{\cal A}_{\rm NLSM}(\overline{\pmb{\sigma}'}_n)={\cal T}^\epsilon[\overline{\pmb{\sigma}'}_n]\Big({\cal L}^{\W\epsilon}\,{\cal T}^{\W\epsilon}[\overline{a,b}]\Big){\cal A}^{\epsilon,\W\epsilon}_{\rm GR}(\pmb{H}_n)\,,
\eea
one can get
\bea
& &{\cal A}_{\rm NLSM}(\overline{\pmb{\sigma}'}_n)={\cal T}^\epsilon[\overline{\pmb{\sigma}'}_n]{\cal A}^\epsilon_{\rm BI}(\pmb{P}_n)\,.
\eea
Thus, the full unified web for tree amplitudes of different theories is involved in Table \ref{tab:unifying}.

\section{From GR to YM and BAS}
\label{GR-YM-BAS}

In this section, we discuss the differential operator which links the $1$-loop Feynman integrands of GR, YM and BAS together.
Based on the structure of CHY integrands of these theories in Table \ref{tab:1-loop-theories}, it is sufficient to find the operator which transmutes the building block ${\cal F}\,{\bf Pf}' \Psi$ to the $1$-loop Parke-Taylor fator
$PT(\overline{+,\sigma_1,\cdots,\sigma_n,-})$, then sum over these operators cyclicly to get $PT_\circ(\overline{\sigma_1,\cdots,\sigma_n})$. In subsection \ref{construction-insertion}, we construct the operator ${\cal T}^\epsilon_\circ[\overline{+,\sigma_1,\cdots,\sigma_n,-}]$ satisfies
\bea
{\cal T}^\epsilon_\circ[\overline{+,\sigma_1,\cdots,\sigma_n,-}]\,{\cal F}\,{\bf Pf}'\Psi={\cal F}\,{\cal T}^\epsilon[\overline{+,\sigma_1,\cdots,\sigma_n,-}]\,{\bf Pf}'\Psi\,,
\eea
by general consideration, without respecting to the formula of obtained object at each step.
Then, in subsection \ref{transmute Pf}, we verify that the operator ${\cal T}^\epsilon_\circ[\overline{+,\sigma_1,\cdots,\sigma_n,-}]$ does transmute ${\cal F}\,{\bf Pf}' \Psi$ to the desired object
$PT(\overline{+,\sigma_1,\cdots,\sigma_n,-})$. The physical interpretation of the insertion operators indicates the freedom of choosing the formula of ${\cal T}^\epsilon_\circ[\overline{+,\sigma_1,\cdots,\sigma_n,-}]$, this issue will be discussed in subsection \ref{freedom-inser}. By applying the operator ${\cal T}^\epsilon_{\circ C}[\overline{+,\sigma_1,\cdots,\sigma_n,-}]$ defined as summing over ${\cal T}^\epsilon_\circ[\overline{+,\sigma_1,\cdots,\sigma_n,-}]$ cyclicly, the BAS Feynman integrand can be generated from the YM Feynman integrand, and the YM Feynman integrand can be generated
from the GR Feynman integrand, as can be seen in subsection \ref{relation-gr-ym-bas}.

\subsection{Constructing operator}
\label{construction-insertion}

Let us try to seek the operator which links the $1$-loop GR, YM and BAS Feynman integrands together.
As pointed out before, transmuting these integrands is equivalent to transmuting ${\cal F}\,{\bf Pf}' \Psi$
to the $1$-loop Parke-Taylor factor $PT_\circ(\overline{\sigma_1,\cdots,\sigma_n})$.
Since $PT_\circ(\overline{\sigma_1,\cdots,\sigma_n})$ can be expanded as
\bea
PT_\circ(\overline{\sigma_1,\cdots,\sigma_n})=\sum_{i\in\{1,\cdots,n\}}\,PT(\overline{+,\sigma_i,\cdots,\sigma_{i-1},-})\,,~~~~\label{exp-PT}
\eea
we can construct the operator ${\cal T}^\epsilon_\circ[\overline{+,\sigma_1,\cdots,\sigma_n,-}]$ which transmutes ${\cal F}\,{\bf Pf}' \Psi$ to the tree Parke-Taylor factor
$PT(\overline{+,\sigma_1,\cdots,\sigma_n,-})$, then sum over the operators ${\cal T}^\epsilon_\circ[\overline{+,\sigma_i,\cdots,\sigma_{i-1},-}]$ cyclicly.

Before starting, we point out that if the action of the differential operator at tree level is commutative with taking forward limit, one can conclude that the desired operator
at $1$-loop level is totally the same, since the $1$-loop integrand is obtained via the forward limit. Unfortunately, as can be seen soon, such commutativity is not satisfied.
Thus, the differential operator which links the tree level GR, YM and BAS amplitudes together can not be applied  directly to the $1$-loop case.

Although the tree level operator does not make sense at the $1$-loop level directly, we still hope the effects of desired $1$-loop operator are paralleled to those at tree level, at each step.
More explicitly, we expect the operator ${\cal T}^\epsilon_\circ[\overline{+,\sigma_1,\cdots,\sigma_m,-}]$ to satisfy
\bea
{\cal T}^\epsilon_\circ[\overline{+,\sigma_1,\cdots,\sigma_m,-}]\,{\cal F}\,{\bf Pf}' \Psi={\cal F}\,{\cal T}^\epsilon[\overline{+,\sigma_1,\cdots,\sigma_m,-}]\,{\bf Pf}' \Psi\,,~~~~\label{key-T}
\eea
for each $m$ with $0\leq m\leq n$, where the tree level operator ${\cal T}^\epsilon[\overline{+,\sigma_1,\cdots,\sigma_m,-}]$ is defined in \eref{defin-T}.
At the tree level, the first step of applying ${\cal T}^\epsilon[\overline{+,\sigma_1,\cdots,\sigma_m,-}]$ is to perform the operator ${\cal T}^\epsilon[\overline{a,b}]\equiv\partial_{\epsilon_a\cdot\epsilon_b}$.
This manipulation has two effects, one is reducing the spins of external particles $a$ and $b$ by $1$, another one is creating
the color ordering $\overline{a,b}$, where the legs $a$ and $b$ play the role of reference legs for inserting more legs into the color ordering.
As the analog, our first step should be choosing two reference legs in the color ordering $\overline{+,\sigma_1,\cdots,\sigma_n,-}$, and reduce the corresponding spins simultaneously. Since the $1$-loop Parke-Taylor factor can be expanded as in \eref{exp-PT},
it is nature to choose $+$ and $-$ as reference legs\footnote{After taking the forward limit, $+$ and $-$ are no longer
external legs of the amplitude. But when discussing the Parke-Taylor factor $PT(\overline{+,\sigma_i,\cdots,\sigma_{i-1},-})$, we still call them "leg".}. However, when taking the forward limit, the polarization vectors of tree level external legs $+$ and $-$ are summed as
\bea
\sum_{r}\,(\epsilon_{+}^r)_\mu(\epsilon_{-}^r)_\nu({\cal A}^{\rm tree}_{n+2})^{\mu\nu}=
\sum_{r}\,(\epsilon_{+}^r)_\mu(\epsilon_{+}^r)^\dagger_\nu({\cal A}^{\rm tree}_{n+2})^{\mu\nu}\,,~~~~~\label{sum-polar}
\eea
thus neither $\epsilon_+$ nor $\epsilon_-$ will appear in the $1$-loop integrand. To handle this, we observe that $\sum_r\epsilon^r_+\cdot(\epsilon^r_+)^\dagger=D-2$, thus dividing $\epsilon_+\cdot\epsilon_-$ at tree level is equivalent to dividing $(D-2)$ at $1$-loop level,
due to the summation \eref{sum-polar}.
Here we think the Lorentz vectors as follows, the momenta in $\{k_1,\cdots,k_n,\ell\}$ and polarization vectors in $\{\epsilon_1,\cdots,\epsilon_n\}$ lie in the $d$ dimensional space where $d$ is regarded as a constant, while the polarization vectors $\epsilon_+$ and $\epsilon_-$ are in the $D$ dimensional space where $D$
is regarded as a variable. We can set $D=d$ finally to obtain a physically acceptable object. The reason for this treatment will be seen soon.
The above observation is not enough, due to another effect of $\partial_{\epsilon_+\cdot\epsilon_-}$ at tree level. Since each polarization vector appears once and only once in each term of the amplitude, the operator $\partial_{\epsilon_+\cdot\epsilon_-}$ turns all $\epsilon_+\cdot V$ and $\epsilon_-\cdot V$ except $\epsilon_i\cdot \epsilon_j$ to $0$ at tree level. Here $V$ denotes Lorentz vectors including both polarization vectors and external momenta. Thus, to realize all effects of the operator $\partial_{\epsilon_+\cdot\epsilon_-}$, the associated manipulation at the $1$-loop level can be chosen as
\bea
{\cal D}\,{\cal F}\,{\bf Pf}'\Psi={\partial\over\partial D}\,\Big({\cal F}\,{\bf Pf}'\Psi\Big)\,.~~~~\label{defin-D}
\eea
The reason we regard $D$ as a variable is to make the operator $\partial_D$ to be well defined. The operator ${\cal D}$ defined above selects the terms contain the factor $(D-2)$ in ${\cal F}\,{\bf Pf}'\Psi$,
and annihilates all the other terms. After applying this operator, the obtained object does not depend on $\epsilon_+$ and $\epsilon_-$ anymore,
all remaining Lorentz vectors in ${\cal D}\,{\cal F}\,{\bf Pf}'\Psi$ are $d$ dimensional, as they should be. Thus, we now have
\bea
{\cal D}\,{\cal F}\,{\bf Pf}'\Psi={\cal F}\,{\cal T}^\epsilon[\overline{+,-}]\,{\bf Pf}'\Psi\,,
\eea
at the first step.

With two reference legs $+$ and $-$ on hand, now we need to insert a leg between them, and decrease the spin of the corresponding particle by $1$. Suppose we insert the leg $n$ at this step,
at tree level, the insertion is realized by applying the operator
\bea
{\cal I}^\epsilon_{+n-}\equiv\partial_{\epsilon_n\cdot k_+}-\partial_{\epsilon_n\cdot k_-}\,.~~~~\label{inser-+n-}
\eea
At $1$-loop level, if we apply the operator ${\cal I}^\epsilon_{+n-}$ directly, we will encounter the ambiguity that $\partial_{\epsilon_n\cdot k_+}$ acts not only on $\epsilon_n\cdot k_+$, but also on $\epsilon_n\cdot k_-$, and similar does $\partial_{\epsilon_n\cdot k_-}$, since $k_+=-k_-=\ell$.
To handle this, we observe that in the limit $k_\pm\to \pm \ell$ we have
\bea
\partial_{\epsilon_n\cdot k_+}\,(\epsilon_n\cdot k_-)&=&-\partial_{\epsilon_n\cdot k_-}\,(\epsilon_n\cdot k_-)\,,\nn
\partial_{\epsilon_n\cdot k_+}\,(\epsilon_n\cdot k_+)&=&-\partial_{\epsilon_n\cdot k_-}\,(\epsilon_n\cdot k_+)\,.~~~~\label{instead}
\eea
It means the operator $-\partial_{\epsilon_n\cdot k_-}$ is equivalent to the opertor $\partial_{\epsilon_n\cdot k_+}$.
Thus, it is natural to choose
\bea
{\cal I}^\epsilon_{\circ;+n-}\equiv\partial_{\epsilon_n\cdot \ell}\,.~~~~\label{defin-inser+-}
\eea
Under this choice, all the effects arise from $-\partial_{\epsilon_n\cdot k_-}\,(\epsilon_n\cdot k_-)$ at tree level
are replaced by those from $\partial_{\epsilon_n\cdot k_+}\,(\epsilon_n\cdot k_-)$, and the first line in \eref{instead}
ensures that the resulting object will not be alternated.
Consequently, at the second step, we found
\bea
{\cal I}^\epsilon_{\circ;+n-}\,{\cal D}\,{\cal F}\,{\bf Pf}'\Psi={\cal F}\,{\cal I}^\epsilon_{+n-}\,{\cal T}^\epsilon[\overline{+,-}]\,{\bf Pf}'\Psi\,.
\eea

The next step is to insert another leg, for example the leg $(n-1)$, between $+$ and $n$, and decrease the spin of the external particle $(n-1)$.
At tree level, it is realized via the insertion operator
\bea
{\cal I}^\epsilon_{+(n-1)n}\equiv\partial_{\epsilon_{n-1}\cdot k_+}-\partial_{\epsilon_{n-1}\cdot k_n}\,.
\eea
Here we also encounter the ambiguity arise from the fact $k_+=-k_-=\ell$. In practice, this obstacle can be bypassed
by employing the momentum conservation law. Using the momentum conservation, one can always remove one of external momenta in the formula
of the tree amplitude.
Suppose we remove $k_-$ before taking the forward limit, then $k_-$ will not appear in the numerator of the Feynman integrand. Then, we can safely define the insertion operator at $1$-loop level as
\bea
{\cal I}^\epsilon_{\circ;+(n-1)n}\equiv\partial_{\epsilon_{n-1}\cdot \ell}-\partial_{\epsilon_{n-1}\cdot k_n}\,.~~~~\label{defin-inser+n}
\eea
Thus, we now arrive at
\bea
{\cal I}^\epsilon_{\circ;+(n-1)n}\,{\cal I}^\epsilon_{\circ;+n-}\,{\cal D}\,{\cal F}\,{\bf Pf}'\Psi={\cal F}\,{\cal I}^\epsilon_{+(n-1)n}\,{\cal I}^\epsilon_{+n-}\,{\cal T}^\epsilon[\overline{+,-}]\,{\bf Pf}'\Psi\,.
\eea

The above insertion procedure can be performed recursively to insert other legs between $1$ and $(n-1)$, until the full color ordering
is obtained.

Combining all the manipulations together, the desired operator which satisfies the relation \eref{key-T} is  conjectured as follows
\bea
{\cal T}^\epsilon_\circ[\overline{+,\sigma_1,\cdots,\sigma_m,-}]\equiv\Big(\prod_{i=1}^{m-1}\,{\cal I}^\epsilon_{\circ;+\sigma_i(\sigma_i+1)}\Big){\cal I}^\epsilon_{\circ;+\sigma_m-}\,{\cal D}\,,~~~~\label{defin-T-loop}
\eea
where the operators ${\cal D}$, ${\cal I}^\epsilon_{\circ;+\sigma_m-}$ and ${\cal I}^\epsilon_{\circ;+\sigma_i\sigma_{i+1}}$ are defined in \eref{defin-D},
\eref{defin-inser+-} and \eref{defin-inser+n}, respectively. This operator is constructed by imposing the requirement \eref{key-T} at each step,
thus its physical effect is strictly paralleled to the effect of the tree level operator ${\cal T}^\epsilon[\overline{+,\sigma_1,\cdots,\sigma_m,-}]$.
In the next subsection, we will verify that the operator ${\cal T}^\epsilon_\circ[\overline{+,\sigma_1,\cdots,\sigma_n,-}]$ transmute the ingredient ${\cal F}\,{\bf Pf}'\Psi$ to the Parke-Taylor factor
$PT(\overline{+,\sigma_1,\cdots,\sigma_n,-})$, as implied by our argument in the current subsection.

\subsection{Transmuting ${\cal F}\,{\bf Pf}'\Psi$}
\label{transmute Pf}

Now we apply the conjectured operator ${\cal T}^\epsilon_\circ[\overline{+,\sigma_1,\cdots,\sigma_n,-}]$ to the $1$-loop ingredient ${\cal F}\,{\bf Pf}'\Psi$, and verify that
\bea
{\cal T}^\epsilon_\circ[\overline{+,\sigma_1,\cdots,\sigma_n,-}]{\cal F}\,{\bf Pf}'\Psi=PT(\overline{+,\sigma_1,\cdots,\sigma_n,-})\,,
\eea
up to an overall sign. Without lose of generality, we take $\sigma_i=i$ for simplicity in this subsection.

The first step is to apply the operator ${\cal D}$ to ${\cal F}\,{\bf Pf}'\Psi$. As analysed in the previous subsection,
when applying to $\sum_{r}\,(\epsilon_{+}^r)_\mu(\epsilon_{-}^r)_\nu({\cal A}^{\rm tree}_{n+2})^{\mu\nu}$, the operator ${\cal D}$ turns $\sum_r\epsilon^r_+\cdot\epsilon^r_-$ to $1$, and annihilates all terms do not contain $\sum_r\epsilon^r_+\cdot\epsilon^r_-$.
Thus, we have
\bea
{\cal D}\,{\cal F}\,{\bf Pf}'\Psi={\cal D}\,\Big(\sum_r\,{\bf Pf}'\Psi^r_\ell\Big)={\bf Pf}'\W\Psi_\ell\,,~~~~\label{pfa-step1}
\eea
where the new matrix $\W\Psi_\ell$ is obtained from $\Psi_\ell$ via the replacement
\bea
\epsilon_+\cdot\epsilon_-\to1\,,~~~~\epsilon_+\cdot V\to 0\,,~~~~\epsilon_-\cdot V\to0\,.
\eea
Without lose of generality, we assume the rows and columns in the matrix $\Psi_\ell$ are arranged by the order $\{1,\cdots,n,+,-\}$\footnote{This assumption can be realized by moving lows
and columns. Since $(n+i)^{\rm th}$ row and column will be moved simultaneously while moving $i^{\rm th}$ ones,
the possible $-$ sign will not arise.}, then the matrix $\W\Psi_\ell$ becomes
\bea
\W\Psi_\ell = \left(
         \begin{array}{c|c|c}
           ~~A_{(n+2)\times (n+2)}~~ &  C_{(n+2)\times n} & 0 \\
           \hline
           -C^{\rm T}_{n\times (n+2)} & B_{n\times n} & 0\\
           \hline
           0 & 0 & X_{2\times2}\\
         \end{array}
       \right)=\left(
         \begin{array}{c|c}
          [\Psi_\ell]_{n,+,-:n}& 0\\
           \hline
            0 & [X]_2\\
         \end{array}
       \right)\,\,,~~~~~\label{result-M}
\eea
with
\bea
[X]_2=\left(
         \begin{array}{cc}
          0& {1\over z_{+-}}\\
           {1\over z_{-+}} & 0\\
         \end{array}
       \right)\,.
\eea

Notice that the above derivation is ensured by the observation that $D$ in ${\cal F}\,{\bf Pf}'\Psi$ only arise from $\sum_r\epsilon^r_+\cdot\epsilon^r_-$. A simple way to see this fact
is to use the expanded formula of the reduced Pfaffian proposed by Lam \cite{Lam:2016tlk}, expressed as follows,
\bea
{\bf Pf}'\Psi=-2^{n-1}\sum_{P\in S_{n+2}}\,(-)^p{W_IU_J\cdots U_K\over z_p}\,,~~~~~~~~z_p=z_Iz_J\cdots z_K\,.
\eea
Here $I=\{\alpha_i\}$, $J=\{\beta_j\}$, $K=\{\gamma_k\}$ are ordered subsets of $\{+,1,\cdots,n,-\}$, and $z_I=z_{\alpha_1\alpha_2}\cdots z_{\alpha_m\alpha_1}$,
and so does $z_J$, $z_K$.
In this formula, Lorentz vectors are included in the objects
\bea
W_I=\epsilon^r_+\cdot f_{\alpha_1}\cdots f_{\alpha_m}\cdot\epsilon^r_-\,,~~~~~~~~
U_J={\rm Tr}\Big(f_{\beta_1}\cdots f_{\beta_{m'}}\Big)\,,
\eea
where $f^{\mu\nu}_a$ are field strength tensors defined as $f^{\mu\nu}_a\equiv k_a^\mu\epsilon^\nu_a-\epsilon^\mu_ak^\nu_a$.
After taking $k_{\pm}\to{\pm}\ell$ and summing over $r$, only $W_I$ with $\{\alpha_i\}=\emptyset$ contributes $(D-2)$.

The reduced Pfaffian of $\W\Psi_l$ can be calculated directly as
\bea
{\bf Pf}'\W\Psi_l&=&{\bf Pf}'[\Psi_\ell]_{n,+,-:n}\,{\bf Pf}[X]_2\nn
&=&{-1\over z_{+-}}\,{\bf Pf}[\Psi_\ell]_{n:n}\,{\bf Pf}[X]_2\nn
&=&{1\over z_{+-}z_{-+}}{\bf Pf}[\Psi_\ell]_{n:n}\,.~~~~~\label{pfa-D}
\eea
As can be seen, the desired Parke-Taylor factor $PT(\overline{+,-})$ which indicates the color ordering $\overline{+,-}$ has appeared.
Comparing with the effect of applying ${\cal T}^\epsilon(\overline{+,-})$ to ${\bf Pf}'\Psi$, we see that
\bea
{\cal D}\,{\cal F}\,{\bf Pf}'\Psi={\cal F}\,{\cal T}^\epsilon[\overline{+,-}]\,{\bf Pf}'\Psi\,,
\eea
as argued in the previous subsection.

The next step is to perform the operator ${\cal I}^\epsilon_{\circ;+n-}\equiv \partial_{\epsilon_n\cdot \ell}$. To do this, we observe that $\epsilon_n\cdot \ell$ only occurs in $C_{nn}$ in the matrix $[\Psi_\ell]_{n:n}$. Using the definition of $C_{nn}$, we get
\bea
{\cal I}^\epsilon_{\circ;+n-}\,C_{nn}&=&-{1\over z_{+n}}+{1\over z_{-n}}={z_{-+}\over z_{+n}z_{n-}}\,.
\eea
Then we need to reorganize the remaining part in ${\bf Pf}[\Psi_\ell]_{n:n}$. Using the definition of Pfaffian in \eref{pfa}, we can expand ${\bf Pf}[\Psi_\ell]_{n:n}$ as
\bea
{\bf Pf}[\Psi_\ell]_{n:n}=\sum_{\a\in\Pi}{\bf sgn}(\sigma_{\a})[\Psi_\ell]_{a_1b_1}[\Psi_\ell]_{a_2b_2}\cdots[\Psi_\ell]_{a_nb_n}\,,~~~~\label{pfa-insertion}
\eea
where the element $[\Psi_\ell]_{a_ib_i}$ is the element at the $a_i^{\rm th}$ row and $b_i^{\rm th}$ column in the matrix $[\Psi_\ell]_{n:n}$.
The operator $\partial_{\epsilon_n\cdot l}$ selects terms containing the element $[\Psi_\ell]_{n,2n}$, since $C_{nn}$ is located at the $n^{\rm th}$ row and $2n^{\rm th}$ column. The remaining part after removing $[\Psi_\ell]_{n,2n}$ corresponds to a partition of the
the  set $\{1,2,\cdots,2n\}\setminus\{n,2n\}$, which has the length $2(n-1)$. Such a term appears in  ${\bf Pf}[\Psi_\ell]_{n-1: n-1}$,
weighted by a new signature ${\bf sgn}(\sigma_{\W\a})$, where  the
new matrix $[\Psi_\ell]_{n-1: n-1}$ is obtained from the original one $[\Psi_\ell]_{n: n}$ by deleting $n^{\rm th}$ and $2n^{\rm th}$ rows and columns,
and ${\bf sgn}(\sigma_{\W\a})$ corresponds to the partition of the
length-$2(n-1)$ set. By comparing these two special partitions, where one belongs to the original
matrix and one belongs to the new one,
\bea
\a&=&\{(a_1,b_1),(a_2,b_2),\cdots,(n,2n),\cdots,(a_n,b_n)\},\nn
\W\a&=&\{(a_1,b_1),(a_2,b_2),\cdots,(a_n,b_n)\}\,,
\eea
one can get  ${\bf sgn}(\sigma_{\W\a})=(-)^{n-1}{\bf sgn}(\sigma_{\a})$.
Consequently, summing over all survived terms gives
\bea
{\cal I}^\epsilon_{\circ;+n-}\,{\bf Pf}[\Psi_\ell]_{n: n}={(-)^{n-1}z_{-+}\over z_{+n}z_{n-}}{\bf Pf}[\Psi_\ell]_{n-1: n-1}\,.~~~~\label{pfa-iner-n}
\eea
Combining \eref{pfa-step1}, \eref{pfa-D} and \eref{pfa-iner-n} together gives
\bea
{\cal I}^\epsilon_{\circ;+n-}\,{\cal D}\,{\cal F}\,{\bf Pf}'\Psi={(-)^n\over z_{+n}z_{n-}z_{-+}}{\bf Pf}[\Psi_\ell]_{n-1: n-1}\,.~~~~\label{pfa-step2}
\eea
We see that the operator ${\cal I}^\epsilon_{\circ;+n-}$ transmutes the Parke-Taylor factor $PT(\overline{+,-})$ to the new one $PT(\overline{+,n,-})$ which indicates the color
ordering $\overline{+,n,-}$. At this step we have
\bea
{\cal T}^\epsilon_{\circ}[\overline{+,n,-}]\,{\cal F}\,{\bf Pf}'\Psi={\cal F}\,{\cal T}^\epsilon[\overline{+,n,-}]\,{\bf Pf}'\Psi\,,
\eea
which validates our main idea.

Then we need to apply the operator ${\cal I}^\epsilon_{\circ;+(n-1)n}\equiv\partial_{\epsilon_{n-1}\cdot \ell}-\partial_{\epsilon_{n-1}\cdot k_n}$.
Notice that in the matrix $[\Psi_\ell]_{n-1: n-1}$ the Lorentz invariants $\epsilon_{n-1}\cdot \ell$ and $\epsilon_{n-1}\cdot k_n$ only
appear in the element $C_{(n-1)(n-1)}$. To avoid the ambiguity that the operator $\partial_{\epsilon_{n-1}\cdot \ell}$ acts on $\epsilon_{n-1}\cdot k_-$, we use the momentum conservation law and the gauge condition $\epsilon_i\cdot k_i$ to rewrite the elements $C_{ii}$ as
\bea
C_{ii}=\sum_{j\neq i,-}\,{(k_j\cdot\epsilon_i)z_{-j}\over z_{ji}z_{i-}}\,.~~~~\label{rewrite-C}
\eea
In the rewritten form, the momenta $k_-=-\ell$ has been removed. A little algebra yields
\bea
{\cal I}^\epsilon_{\circ;+(n-1)n}\,C_{(n-1)(n-1)}&=&{z_{+n}\over z_{+(n-1)}z_{n(n-1)}}\,.
\eea
Using this, together with the discussion about partitions in the previous step, we find
\bea
{\cal I}^\epsilon_{\circ;+(n-1)n}\,{\bf Pf}[\Psi_\ell]_{n-1: n-1}={(-)^{n-2}z_{+n}\over z_{+(n-1)}z_{n(n-1)}}{\bf Pf}[\Psi_\ell]_{n-2:n-2}\,,~~~~\label{pfa-iner-n-1}
\eea
thus
\bea{\cal I}^\epsilon_{\circ;+(n-1)n}
\,{\cal I}^\epsilon_{\circ;+n-}\,{\cal D}\,{\cal F}\,{\bf Pf}'\Psi={(-)^n(-)^{n-1}\over z_{+(n-1)}z_{(n-1)n}z_{n-}z_{-+}}{\bf Pf}[\Psi_\ell]_{n-2: n-2}\,.~~~~\label{pfa-step3}
\eea
Here the matrix $[\Psi_\ell]_{n-2: n-2}$ is obtained from $[\Psi_\ell]_{n-1: n-1}$ by deleting $(n-1)^{\rm th}$ and $2(n-1)^{\rm th}$ rows and columns.
At this step, the Parke-Taylor factor $PT(\overline{+,(n-1),n,-})$ has been generated. We again notice that
\bea
{\cal T}^\epsilon_{\circ}[\overline{+,n-1,n,-}]\,{\cal F}\,{\bf Pf}'\Psi={\cal F}\,{\cal T}^\epsilon[\overline{+,n-1,n,-}]\,{\bf Pf}'\Psi\,.
\eea
The manipulation at this step can be performed recursively, and finally we arrive at
\bea
{\cal T}^\epsilon_\circ[\overline{+,1,\cdots,n,-}]\,{\cal F}\,{\bf Pf}'\Psi=(-)^{n(n+1)\over2}PT(\overline{+,1,\cdots,n,-})\,.
\eea

Thus, we conclude that the operator ${\cal T}^\epsilon_\circ[\overline{+,\sigma_1,\cdots,\sigma_n,-}]$ transmutes ${\cal F}\,{\bf Pf}'\Psi$ to the tree Parke-Taylor factor $PT(\overline{+,\sigma_1,\cdots,\sigma_n,-})$, up to an overall sign $(-)^{n(n+1)\over2}$. We also verified that
the key requirement \eref{key-T} is satisfied at each step.

\subsection{Freedom of choosing insertion operators}
\label{freedom-inser}

In the previous subsection, the color ordering $\overline{+,1,\cdots,n,-}$ is obtained by giving two reference legs $+$ and $-$
at the first step, then inserting the leg $n$ between $+$ and $-$, and then inseting $(n-1)$ between $+$ and $n$, and so on.
It is natural to ask if the insertions can be done in different manners. For example, consider the color ordering
$\overline{+,1,2,3,-}$. The algorithm in the previous subsection is as follows:
\begin{itemize}
\item Create reference legs $+$ and $-$;
\item Insert $3$ between $+$ and $-$;
\item Insert $2$ between $+$ and $3$;
\item Insert $1$ between $+$ and $2$.
\end{itemize}
However, it is natural to image other manners such as
\begin{itemize}
\item Create reference legs $+$ and $-$;
\item Insert $2$ between $+$ and $-$;
\item Insert $1$ between $+$ and $2$;
\item Insert $3$ between $2$ and $-$.
\end{itemize}

Let us verify the above alternative manner by rigorous calculation.
The definitions of insertion operators ${\cal I}^\epsilon_{\circ;+2-}$ and ${\cal I}^\epsilon_{\circ;+12}$
can be found in \eref{defin-inser+-} and \eref{defin-inser+n}, respectively.
As the analog of the tree level insertion operator, we also introduce the operator
\bea
{\cal I}^\epsilon_{\circ;23-}\equiv\partial_{\epsilon_3\cdot k_2}-\partial_{\epsilon_3\cdot k_-}
=\partial_{\epsilon_3\cdot k_2}+\partial_{\epsilon_3\cdot \ell}\,.
\eea
When applying ${\cal I}^\epsilon_{\circ;23-}$ to $C_{33}$, to avoid the ambiguity that $\partial_{\epsilon_3\cdot \ell}$
acts on $\epsilon_3\cdot k_+$, one should rewrite $C_{33}$ as
\bea
C_{33}=\sum_{j\neq 3,-}\,{(k_j\cdot\epsilon_3)z_{+j}\over z_{j3}z_{3+}}\,.
\eea
Then, following the calculation in the previous subsection, we find that the alternative operator, defined by
\bea
{\cal T}^\epsilon_\circ[\overline{+,1,2,3,-}]\equiv{\cal I}^\epsilon_{\circ;23-}\,{\cal I}^\epsilon_{\circ;+12}\,{\cal I}^\epsilon_{\circ;+2-}\,{\cal D}\,,
\eea
transmutes ${\cal F}\,{\bf Pf}'\Psi$ to the Parke-Taylor factor $PT(\overline{+,1,2,3,-})$, as expected.

It is straightforward to generalized the above discussion to the general case with arbitrary number of external particles. Thus, to get the correct
Parke-Taylor factor $PT(\overline{+,1,\cdots,n,-})$, the corresponding operator ${\cal T}^\epsilon_\circ[\overline{+,1,\cdots,n,-}]$ has the freedom of choosing
insertion operators. The insertion operators can be separated into $4$ classes, defined as follows,
\bea
& &{\cal I}^\epsilon_{\circ;+i-}\equiv\partial_{\epsilon_i\cdot \ell}\,,~~~~~~~~~~~~~~~~~~
{\cal I}^\epsilon_{\circ;+kj}\equiv\partial_{\epsilon_k\cdot \ell}-\partial_{\epsilon_k\cdot k_j}\,,\nn
& &{\cal I}^\epsilon_{\circ;ik-}\equiv\partial_{\epsilon_k\cdot k_i}+\partial_{\epsilon_k\cdot \ell}\,,~~~~~~~~
{\cal I}^\epsilon_{\circ;ikj}\equiv\partial_{\epsilon_k\cdot k_i}-\partial_{\epsilon_k\cdot k_j}\,,
\eea
each ${\cal I}^\epsilon_{\circ;abc}$ can be understood as turning the Parke-Taylor factor $PT(\overline{\cdots,a,c,\cdots})$ to $PT(\overline{\cdots,a,b,c,\cdots})$.
All ${\cal T}^\epsilon_\circ[\overline{+,\sigma_1,\cdots,\sigma_n,-}]$ constructed via the above interpretation satisfy our requirement.

\subsection{Relations among GR, YM and BAS Feynman integrands}
\label{relation-gr-ym-bas}

We have constructed the operator ${\cal T}^\epsilon_\circ[\overline{+,\sigma_1,\cdots,\sigma_n,-}]$ which transmutes ${\cal F}\,{\bf Pf}'\Psi$ to
$PT(\overline{+,\sigma_1,\cdots,\sigma_n,-})$.
Now we introduce the operator ${\cal T}^\epsilon_{\circ C}[\overline{\sigma_1,\cdots,\sigma_n}]$ as
\bea
{\cal T}^\epsilon_{\circ C}[\overline{\sigma_1,\cdots,\sigma_n}]\equiv\sum_{i=1}^n\,{\cal T}^\epsilon_\circ[\overline{+,\sigma_i,\cdots,\sigma_{i-1},-}]\,.
\eea
Clearly, the operator ${\cal T}^\epsilon_{\circ C}[\overline{\sigma_1,\cdots,\sigma_n}]$ defined above transmutes ${\cal F}\,{\bf Pf}'\Psi$ to
$PT_{\circ}(\overline{\sigma_1,\cdots,\sigma_n})$.
It is natural to expect that this operator transmutes the $1$-loop GR CHY integrand to the $1$-loop YM CHY integrand, and transmutes the $1$-loop YM CHY integrand to the BAS CHY integrand, due to the structures of $1$-loop CHY integrands established in Table \ref{tab:1-loop-theories}. Since this operator does not modify the measure of CHY contour integral, it is commutative with the CHY contour integral. Thus, we conclude this operator transmutes the $1$-loop GR Feynman integrand to the $1$-loop YM Feynman integrand, and transmutes
the $1$-loop YM Feynman integrand to the BAS Feynman integrand.

To really achieve the goal described above, we require that the polarization vectors $\epsilon_+$ and $\epsilon_-$ are $D$ dimensional, while the polarization vectors $\W\epsilon_+$ and $\W\epsilon_-$ are
$\W D$ dimensional. Now we introduce
\bea
{\cal D}\,{\cal I}_{\circ{\rm GR}}^L={\partial\over \partial D}\,\Big({\cal I}_{\circ{\rm GR}}^L\Big)\,,~~~~~~~~\W{\cal D}\,{\cal I}_{\circ{\rm GR}}^R={\partial\over\partial\W D}\,\Big({\cal I}_{\circ{\rm GR}}^R\Big)\,.
\eea
Here $D$ and $\W D$ are two different variables, thus the operator ${\cal D}$ will not affect ${\cal I}_{\circ{\rm GR}}^R$, while the operator $\W{\cal D}$ will not affect ${\cal I}_{\circ{\rm GR}}^L$.
Then we define two operators
\bea
& &{\cal T}^\epsilon_\circ[\overline{+,\sigma_1,\cdots,\sigma_n,-}]\equiv\Big(\prod_{i=1}^{n-1}\,{\cal I}^\epsilon_{\circ;+\sigma_i\sigma_{i+1}}\Big){\cal I}^\epsilon_{\circ;+\sigma_n-}\,{\cal D}\,,\nn
& &{\cal T}^{\W\epsilon}_\circ[\overline{+,\sigma'_1,\cdots,\sigma'_n,-}]\equiv\Big(\prod_{i=1}^{n-1}\,{\cal I}^{\W\epsilon}_{\circ;+\sigma'_i\sigma'_{i+1}}\Big){\cal I}^{\W\epsilon}_{\circ;+\sigma'_n-}\,\W{\cal D}\,,
\eea
where the insertion operators ${\cal I}^\epsilon_{\circ;abc}$ and ${\cal I}^{\W\epsilon}_{\circ;abc}$ are defined via polarization vectors in $\{\epsilon_i\}$ and $\{\W\epsilon_i\}$, respectively. Applying these two operators, we have
\bea
& &{\cal T}^\epsilon_\circ[\overline{+,\sigma_1,\cdots,\sigma_n,-}]\,\Big({\cal I}_{\circ{\rm GR}}^L\,{\cal I}_{\circ{\rm GR}}^R\Big)=PT(\overline{+,\sigma_1,\cdots,\sigma_n,-})\,{\cal I}_{\circ{\rm GR}}^R\,,\nn
& &{\cal T}^{\W\epsilon}_\circ[\overline{+,\sigma'_1,\cdots,\sigma'_n,-}]\,\Big({\cal I}_{\circ{\rm GR}}^L\,{\cal I}_{\circ{\rm GR}}^R\Big)={\cal I}_{\circ{\rm GR}}^L\,PT(\overline{+,\sigma'_1,\cdots,\sigma'_n,-})\,,
\eea
without any ambiguity. Then, we define
\bea
& &{\cal T}^\epsilon_{\circ C}[\overline{\sigma_1,\cdots,\sigma_n}]\equiv\sum_{i=1}^n\,{\cal T}^\epsilon_\circ[\overline{+,\sigma_i,\cdots,\sigma_{i-1},-}]\,,\nn
& &{\cal T}^{\W\epsilon}_{\circ C}[\overline{\sigma'_1,\cdots,\sigma'_n}]\equiv\sum_{i=1}^n\,{\cal T}^{\W\epsilon}_\circ[\overline{+,\sigma'_i,\cdots,\sigma'_{i-1},-}]\,.
\eea
Applying the above operators ${\cal T}^\epsilon_{\circ C}[\sigma_1,\cdots,\sigma_n]$ and ${\cal T}^{\W\epsilon}_{\circ C}[\overline{\sigma'_1,\cdots,\sigma'_n}]$
to $\Big({\cal I}_{\circ{\rm GR}}^L\,{\cal I}_{\circ{\rm GR}}^R\Big)$, and using the commutativity between the operators and the CHY integral,
we arrive at the following relations
\bea
{\bf I}^\epsilon_{\rm YM}(\overline{\sigma'_1,\cdots,\sigma'_n})&=&{\cal T}^{\W\epsilon}_{\circ C}[\overline{\sigma'_1,\cdots,\sigma'_n}]\,{\bf I}^{\epsilon,\W\epsilon}_{\rm GR}({\pmb H}_n)\,,\nn
{\bf I}^{\W\epsilon}_{\rm YM}(\overline{\sigma_1,\cdots,\sigma_n})&=&{\cal T}^{\epsilon}_{\circ C}[\overline{\sigma_1,\cdots,\sigma_n}]\,{\bf I}^{\epsilon,\W\epsilon}_{\rm GR}({\pmb H}_n)\,,\nn
{\bf I}_{\rm BAS}(\overline{\sigma_1,\cdots,\sigma_n}|\overline{\sigma'_1,\cdots,\sigma'_n})&=&{\cal T}^{\epsilon}_{\circ C}[\overline{\sigma_1,\cdots,\sigma_n}]\,{\bf I}^{\epsilon}_{\rm YM}(\overline{\sigma'_1,\cdots,\sigma'_n})\nn
&=&{\cal T}^{\W\epsilon}_{\circ C}[\overline{\sigma'_1,\cdots,\sigma'_n}]\,{\bf I}^{\W\epsilon}_{\rm YM}(\overline{\sigma_1,\cdots,\sigma_n})\nn
&=&{\cal T}^{\epsilon}_{\circ C}[\overline{\sigma_1,\cdots,\sigma_n}]\,
{\cal T}^{\W\epsilon}_{\circ C}[\overline{\sigma'_1,\cdots,\sigma'_n}]\,{\bf I}^{\epsilon,\W\epsilon}_{\rm GR}({\pmb H}_n)\,.
\eea
From these relations, we see that by applying operators ${\cal T}^{\epsilon}_{\circ C}[\overline{\sigma_1,\cdots,\sigma_n}]$ and ${\cal T}^{\W\epsilon}_{\circ C}[\overline{\sigma_1,\cdots,\sigma_n}]$, the $1$-loop YM Feynman integrand can be generated from the GR Feynman integrand, and
the BAS Feynman integrand can be generated from the YM Feynman integrand.

Some remarks are in order. When applying to ${\bf I}^{\epsilon,\W\epsilon}_{\rm GR}({\pmb H}_n)$, two operators ${\cal T}^{\W\epsilon}_{\circ C}[\overline{\sigma'_1,\cdots,\sigma'_n}]$ and ${\cal T}^{\epsilon}_{\circ C}[\overline{\sigma_1,\cdots,\sigma_n}]$ work well as long as ${\bf I}^{\epsilon,\W\epsilon}_{\rm GR}({\pmb H}_n)$ includes two parameters $D$ and $\W D$. Of course we have $D=\W D=d$, but we need to separate $d$
into two sectors and denote them by $D$ and $\W D$ respectively. This requirement is realized naturally via the forward limit method, since $D$
and $\W D$
arise from two distinguished sectors of polarization vectors. Thus, to apply the operators considered above, one need to denote $d-2=\sum_r\epsilon_+^r\cdot\epsilon_-^r$ as $(D-2)$ and denote $d-2=\sum_r\W\epsilon_+^r\cdot\W\epsilon_-^r$ as $(\W D-2)$ when using the forward limit method to write down the $1$-loop Feynman integrand.

Before ending this subsection, we emphasize that the operators ${\cal T}^{\epsilon}_{\circ C}[\overline{\sigma_1,\cdots,\sigma_n}]$ and ${\cal T}^{\W\epsilon}_{\circ C}[\overline{\sigma_1,\cdots,\sigma_n}]$ are not commutative with the $1$-loop Feynman integral, since they affect on Lorentz
invariants $\epsilon_i\cdot l$ and $\W\epsilon_j\cdot l$ which depend on the loop momentum. Thus, the relations discussed above only hold
at the Feynman integrands level, i.e., they are not satisfied at the $1$-loop amplitudes level.

\section{Other theories}
\label{other theory}

In the previous section, we constructed the operator ${\cal T}^{\epsilon}_{\circ C}[\overline{\sigma_1,\cdots,\sigma_n}]$ which links the $1$-loop Feynman
integrands of GR, YM and BAS together. These relations are inherited from the relations at tree level, and the basic idea is to seek the operator
${\cal O}^\epsilon_\circ$ satisfying ${\cal O}^\epsilon_\circ\,{\cal F}\,{\bf Pf}'\Psi={\cal F}\,{\cal O}^\epsilon\,{\bf Pf}'\Psi$. At tree level, the unified web includes a large variety of theories, as can be seen in Table \ref{tab:unifying}.
Thus, the aim of this section is to apply the same idea to other theories included in Table \ref{tab:unifying}, to construct more $1$-loop level operators which generates the $1$-loop Feynman integrands of other theories.

The CHY integrands for theories under consideration in this section have not been given explicitly in literatures, but can be obtained easily via the forward limit method. For the $1$-loop CHY formula for a new theory, one may encounter the problem of singular solutions of scattering equations. Fortunately, all CHY integrands which will be considered in this section are homogeneous in $\ell^\mu$, as can be verified directly. As discussed in subsection.\ref{forwardlimit}, using the
conclusion in \cite{Cachazo:2015aol}, we can claim that all singular solutions contribute to scaleless integrals which vanish under the dimensional regularization, and ignore them.

In subsection \ref{ssEYM-ssYMS}, we show that the operator ${\cal T}^{\epsilon}_{\circ C}[\overline{\sigma_1,\cdots,\sigma_m}]$ with $m<n$ transmutes
the $1$-loop $n$-point GR Feynman integrand to the single trace $1$-loop $n$-point EYM Feynman integrand, with a gluon running in the loop, and also transmutes
the $1$-loop $n$-point YM Feynman integrand to the single trace $1$-loop $n$-point BAS Feynman integrand, with a scalar running in the loop. We call them the special Feynman integrands of single trace EYM and single trace YMS, and denote them as ssEYM and ssYMS, respectively. In subsection \ref{BI,NLSM,SG}, we discuss the operator which transmutes the $1$-loop GR Feynman integrand to the BI and SG Feynman integrands, and also transmutes the $1$-loop YM Feynman integrand to the NLSM integrand. In subsection \ref{EM-DBI}, we consider the operator which transmutes $1$-loop
GR Feynman integrand to the EM Feynman integrand, and transmutes the $1$-loop BI Feynman integrand to the DBI Feynman integrand. In this subsection,
new situation arises from the fact the virtual particle running in the loop is not unique, which means the operator ${\cal F}$ is not unique, thus one can not expect ${\cal O}_\circ\,{\cal F}\,\bullet={\cal F}\,{\cal O}\,\bullet$ for an individual ${\cal F}$. We will show how to generalize our idea to this new situation.

\subsection{ssEYM and ssYMS}
\label{ssEYM-ssYMS}

In the previous section, we constructed the operator ${\cal T}^{\epsilon}_\circ[\overline{+,\sigma_1,\cdots,\sigma_m,-}]$ satisfying
the relation \eref{key-T}.
We also proved that the operator ${\cal T}^{\epsilon}_\circ[\overline{+,\sigma_1,\cdots,\sigma_n,-}]$ transmutes
the object ${\cal F}\,{\bf Pf}'\Psi$ to the Parke-Taylor factor $PT(\overline{+,\sigma_1,\cdots,\sigma_n,-})$, which contributes to the $1$-loop CHY integrands of YM and BAS. It is natural to ask the physical interpretation of the resulting object after performing ${\cal T}^{\epsilon}_\circ[\overline{+,\sigma_1,\cdots,\sigma_m,-}]$,
with $0\leq m<n$.

We begin the discussion of this subject by applying
${\cal T}^\epsilon_\circ[+,\sigma_1,\cdots,\sigma_m,-]$ to the $1$-loop GR Feynman integrand.
At tree level, the corresponding operator
${\cal T}^\epsilon[+,\sigma_1,\cdots,\sigma_m,-]$
transmutes the GR amplitude to the single trace EYM one as follows
\bea
{\cal T}^\epsilon[+,\sigma_1,\cdots,\sigma_m,-]\,{\cal A}^{\epsilon,\W\epsilon}_{\rm GR}({\pmb H}_{n+2})={\cal A}^{\epsilon,\W\epsilon}_{\rm EYM}(\overline{+,\sigma_1,\cdots,\sigma_m,-};{\pmb{H}}_{n-m})\,.
\eea
Using the relation \eref{key-T}, we conclude that the object
${\cal T}^\epsilon_\circ[\overline{+,\sigma_1,\cdots,\sigma_m,-}]\,{\cal F}\,{\cal A}^{\epsilon,\W\epsilon}_{\rm GR}({\pmb H}_{n+2})$
can be understood as the forward limit ${\cal F}\,{\cal A}^{\epsilon,\W\epsilon}_{\rm EYM}(\overline{+,\sigma_1,\cdots,\sigma_m,-};{\pmb{H}}_{n-m})$, thus contributes to the $1$-loop EYM Feynman integrand.

\begin{figure}
  \centering
  \includegraphics[width=12cm]{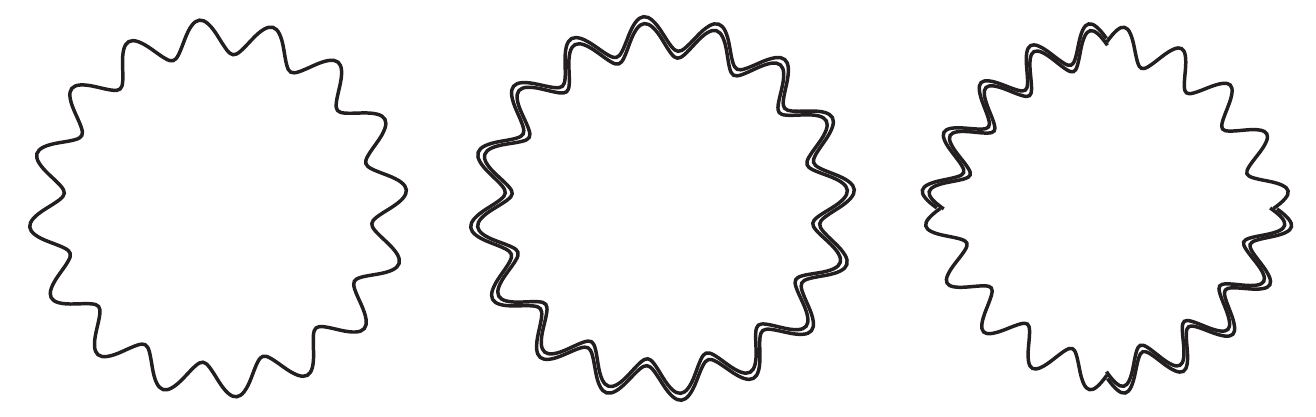} \\
  \caption{Typical structures of virtual particles in the loop, the single wavy lines denote gluons or photons, while the double wavy lines denote gravitons.}\label{virtual}
\end{figure}

\begin{figure}
  \centering
  \includegraphics[width=12cm]{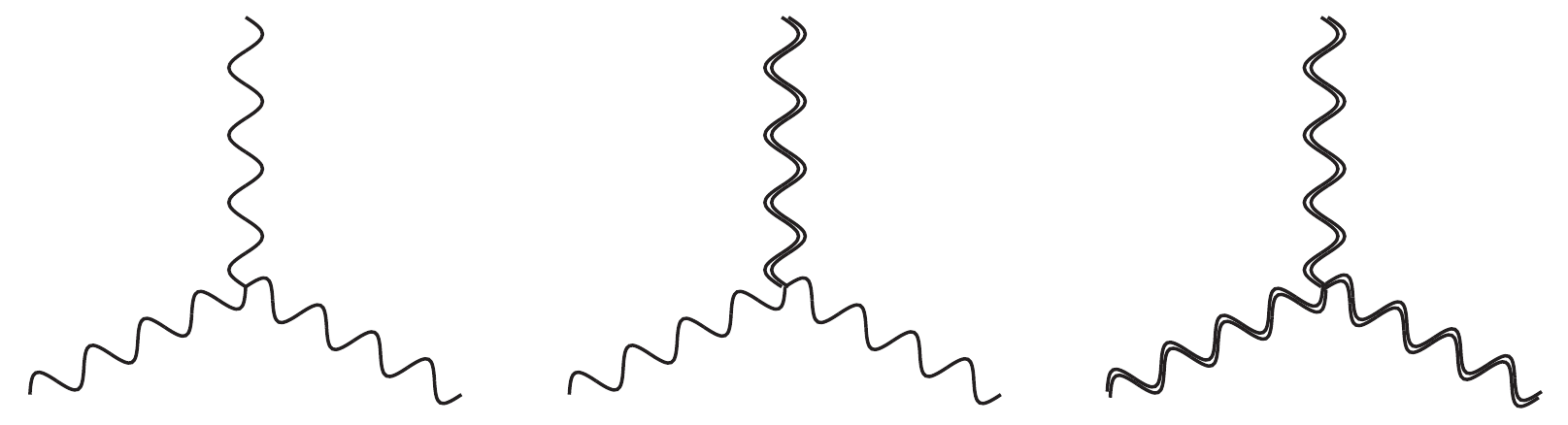} \\
  \caption{Three vertices of EYM theory, the single wavy lines denote gluons while the double wavy lines denote gravitons. In the second diagram, the single wavy lines can also be understood as photons.}\label{vertex}
\end{figure}

However, one can not expect that the full $1$-loop single trace EYM  Feynman integrand with fixed external legs and color ordering can be obtained
through the above manipulation. The reason is, for the EYM theory, the virtual
particle running in the loop can be either gluon or graviton, as shown in Fig.\ref{virtual}, but the operator ${\cal D}$ indicates that the
forward limit is took for two external gluons thus the diagrams with only a graviton running in the loop are excluded.
Now we argue that after performing the operator ${\cal D}$, the only candidate for the virtual particle in the loop is a gluon.
The EYM theory includes three interaction vertices in Fig.\ref{vertex}, these vertices indicate that for the tree EYM amplitude including only two external gluons, one can always start from one external
gluon, go along the gluon lines, and arrive at another one. It means, after taking the forward limit for two external gluons,
a closed loop contains only gluon lines is obtained. At the $1$-loop level, this observation is sufficient to fix the virtual particle as a gluon. Thus, the operator ${\cal D}$ turns the
internal graviton running in the loop to a gluon. Consequently, after applying  ${\cal T}^\epsilon_\circ[\overline{+,-}]={\cal D}$ to the $1$-loop GR Feynman integrand, the ontained object is nothing but the
$1$-loop EYM Feynman
integrand with a gluon in the loop and all external particles are gravitons.

The calculation in subsection \ref{transmute Pf} shows that the operator ${\cal T}^\epsilon_\circ[\overline{+,\sigma_1,\cdots,\sigma_m,-}]$ transmutes
${\cal F}\,{\bf Pf}'\Psi$ as follows
\bea
{\cal T}^\epsilon_\circ[\overline{+,\sigma_1,\cdots,\sigma_m,-}]\,{\cal F}\,{\bf Pf}'\Psi=PT(\overline{+,\sigma_1,\cdots,\sigma_m,-})\,{\bf Pf}[\Psi_\ell]_{n-m: n-m}\,,
~~~~\label{EYM-st-partial-trace}
\eea
where the matrix $[\Psi_\ell]_{n-m: n-m}$ is obtained from $[\Psi_\ell]_{n:n}$ by deleting rows and columns labeled by $\sigma_1,\cdots,\sigma_m$,
and $\sigma_1+n,\cdots,\sigma_m+n$. The Parke-Taylor factor $PT(\overline{+,\sigma_1,\cdots,\sigma_m,-})$ indicates the color ordering $\overline{\sigma_1,\cdots,\sigma_m}$ at $1$-loop level, thus the r.h.s of \eref{EYM-st-partial-trace} contributes to the $1$-loop single trace color ordered EYM  CHY integrand, with a gluon running in the loop. We emphasize that  taking the forward limit for two external gluons for the tree EYM CHY integrand also gives the r.h.s of \eref{EYM-st-partial-trace}, as can be verified directly. Defining
\bea
{\cal T}^\epsilon_{\circ C}[\overline{\sigma_1,\cdots,\sigma_m}]\equiv\sum_{i=1}^m\,{\cal T}^\epsilon_\circ[\overline{+,\sigma_i,\cdots,\sigma_{i-1},-}]\,,
\eea
we arrive at the relation
\bea
{\cal T}^\epsilon_{\circ C}[\overline{\sigma_1,\cdots,\sigma_m}]\,{\cal F}\,{\bf Pf}'\Psi=PT_{\circ}(\overline{\sigma_1,\cdots,\sigma_m})\,{\bf Pf}[\Psi_\ell]_{n-m: n-m}\,,
~~~~\label{EYM-st-partial1}
\eea
where $PT_{\circ}(\overline{\sigma_1,\cdots,\sigma_m})$ is the $1$-loop Parke-Taylor factor defined in \eref{PT-C}. We immediately get the conclusion that applying ${\cal T}^\epsilon_{\circ C}[\overline{\sigma_1,\cdots,\sigma_m}]$ to the $1$-loop GR Feynman integrand, the obtained result is the single trace $1$-loop EYM Feynman integrand,
with a gluon running in the loop, expressed as follows
\bea
{\cal T}^\epsilon_{\circ C}[\overline{\sigma_1,\cdots,\sigma_m}]\,{\bf I}^{\epsilon,\W\epsilon}_{\rm GR}({\pmb H}_n)={\bf I}^{\epsilon,\W\epsilon}_{\rm EYM}(\overline{\sigma_1,\cdots,\sigma_m};{\pmb H}_{n-m})\,.
\eea
This statement is valid for all lengths-$m$ ordered sets
$\{\overline{\sigma_1,\cdots,\sigma_m}\}$ with $0\leq m< n$, therefore
\bea
{\cal T}^\epsilon_\circ[\overline{+,\cdots,a,b,c,\cdots,-}]\,{\cal F}\,{\cal A}^{\epsilon,\W\epsilon}_{\rm GR}({\pmb H}_n)
&=&{\cal I}^\epsilon_{\circ;abc}\,{\cal T}^\epsilon_\circ[\overline{+,\cdots,a,c,\cdots,-}]\,{\cal F}\,{\cal A}^{\epsilon,\W\epsilon}_{\rm GR}({\pmb H}_n)\nn
&=&{\cal I}^\epsilon_{\circ;abc}{\cal F}\,{\cal A}^{\epsilon,\W\epsilon}_{\rm EYM}(+,\cdots,a,c,\cdots,-;{\pmb H}_{n-m+1})\,\,.
\eea
This equality supports the interpretation of the $1$-loop insertion operator
${\cal I}^\epsilon_{\circ;abc}$: it turns the graviton $b$ to a gluon, and inserts the leg $b$ between $a$ and $c$ in the color ordering.

At the tree level, the YMS amplitude can be generated from the YM one, by applying the operator ${\cal T}^\epsilon[\overline{+,\sigma_1,\cdots,\sigma_m,-}]$, as can be seen in Table.\ref{tab:unifying}.
Repeating the discussion for the EYM case, we conclude that the operator ${\cal T}^\epsilon_{\circ C}[\overline{\sigma_1,\cdots,\sigma_m}]$ transmutes the $1$-loop
color ordered YM Feynman integrand to the $1$-loop double color ordered YMS Feynman integrand, with a scalar running in the loop.

\subsection{BI, NLSM, SG}
\label{BI,NLSM,SG}

In this subsection, we show that the operator ${\cal L}^\epsilon_\circ\,{\cal D}$ links the Feynman integrands for GR, BI, NLSM and
SG theories together. The definition of ${\cal L}^\epsilon_\circ$ is the same as for the tree level case, except taking $k_\pm=\pm \ell$.

In subsection \ref{transmute Pf}, we found that
\bea
{\cal D}\,{\cal F}\,{\bf Pf}'\,\Psi={\cal F}\,{\cal T}^\epsilon[\overline{+,-}]\,{\bf Pf}'\,\Psi\,.~~~~\label{commu-D-F}
\eea
Thus we immediately get
\bea
{\cal L}^\epsilon_\circ\,{\cal D}\,{\cal F}\,{\bf Pf}'\,\Psi={\cal F}\,{\cal L}^\epsilon\,{\cal T}^\epsilon[\overline{+,-}]\,{\bf Pf}'\,\Psi\,,~~~~\label{loop-operator-l}
\eea
since the relation ${\cal L}^\epsilon_\circ{\cal F}\bullet={\cal F}{\cal L}^\epsilon\bullet$ is manifest. At the tree level,
the object
\bea
{\cal L}^\epsilon\,{\cal T}^\epsilon[\overline{+,-}]\,{\bf Pf}'\,\Psi=({\bf Pf}'\,A)^2
\eea
contributes to the BI, NLSM and SG integrands, and we have
\bea
& &{\cal L}^\epsilon\,{\cal T}^\epsilon[\overline{+,-}]\,{\cal A}^{\epsilon,\W\epsilon}_{\rm GR}({\pmb H}_{n+2})={\cal A}^{\W\epsilon}_{\rm BI}({\pmb P}_{n+2})\,,\nn
& &{\cal L}^\epsilon\,{\cal T}^\epsilon[\overline{+,-}]\,{\cal A}^{\epsilon}_{\rm YM}(\overline{+,\sigma_1,\cdots,\sigma_n,-})={\cal A}_{\rm NLSM}(\overline{+,\sigma_1,\cdots,\sigma_n,-})\,,\nn
& &{\cal L}^{\W\epsilon}\,{\cal T}^{\W\epsilon}[\overline{+,-}]\,{\cal A}^{\W\epsilon}_{\rm BI}({\pmb P}_{n+2})={\cal A}_{\rm SG}({\pmb S}_{n+2})\,,~~~~\label{rela-tree}
\eea
as shown in Table.\ref{tab:unifying}.
Taking the forward limit for tree amplitudes gives rise to the $1$-loop Feynman integrands for BI, NLSM and SG, respectively, i.e.,
\bea
& &{1\over \ell^2}\,{\cal F}\,{\cal A}^{\W\epsilon}_{\rm BI}({\pmb P}_{n+2})={\bf I}^{\W\epsilon}_{\rm BI}({\pmb P}_n)\,,\nn
& &{1\over \ell^2}\,\sum_{i\in\{1,\cdots,n\}}\,{\cal F}\,{\cal A}_{\rm NLSM}(\overline{+,\sigma_i,\cdots,\sigma_{i-1},-})={\bf I}_{\rm NLSM}(\overline{\sigma_1,\cdots,\sigma_n})\,,\nn
& &{1\over \ell^2}\,{\cal F}\,{\cal A}_{\rm SG}({\pmb S}_{n+2})={\bf I}_{\rm SG}({\pmb S}_n)\,.~~~~\label{rela-forward}
\eea
Notice that to get the color ordered Feynman integrand ${\bf I}_{\rm NLSM}(\overline{\sigma_1,\cdots,\sigma_n})$, the cyclic summation
of color orderings is required.
Each one of three theories includes only one kind of particles, thus there is no ambiguity about the virtual particle running in
the loop. Combining \eref{loop-operator-l}, \eref{rela-tree} and \eref{rela-forward} together, we obtain the following relations
\bea
& &{\cal L}^\epsilon_\circ\,{\cal D}\,{\bf I}^{\epsilon,\W\epsilon}_{\rm GR}({\pmb H}_n)={\bf I}^{\W\epsilon}_{\rm BI}({\pmb P}_n)\,,\nn
& &{\cal L}^\epsilon_\circ\,{\cal D}\,{\bf I}^{\epsilon}_{\rm YM}(\overline{\sigma_1,\cdots,\sigma_n})={\bf I}_{\rm NLSM}(\overline{\sigma_1,\cdots,\sigma_n})\,,\nn
& &{\cal L}^{\W\epsilon}_\circ\,\W{\cal D}\,{\bf I}^{\W\epsilon}_{\rm BI}({\pmb P}_n)={\bf I}_{\rm SG}({\pmb S}_n)\,.~~~~\label{rela-loop}
\eea

To verify the relations in \eref{rela-loop}, let us apply the operator ${\cal L}^\epsilon_\circ\,{\cal D}$ to ${\cal F}\,{\bf Pf}'\,\Psi$.
In the subsection \ref{transmute Pf}, we found that
\bea
{\cal D}\,{\cal F}\,{\bf Pf}'\,\Psi={1\over z_{+-}}\,{\bf Pf}'\,[\Psi_l]_{n,+,-:n}\,,~~~~\label{eff-D-A}
\eea
where the $2(n+1)\times 2(n+1)$ matrix $[\Psi_l]_{n,+,-:n}$ is obtained from $\Psi_l$ by deleting rows and columns labeled by $\epsilon_+$ and $\epsilon_-$. To continue,
we need to apply ${\cal L}^\epsilon_\circ$ to ${\bf Pf}'\,[\Psi_l]_{n,+,-:n}$. Since each polarization vector appears once and only once in each term of the reduced Pffafian,
the operator $\partial_{\epsilon_i\cdot k_j}$ turns $\epsilon_i\cdot k_j$ to $1$ and annihilates all other $\epsilon_i\cdot V$ simultaneously.
This observation indicates that
\bea
{\cal L}^\epsilon_\circ\,{\bf Pf}'\,[\Psi_l]_{n,+,-:n}={\bf Pf}'\,\W A={-1\over z_{+-}}\,{\bf Pf} \left(
         \begin{array}{c|c}
           ~~A_{n\times n}~~ &  A_{n\times n}\\
           \hline
          A_{n\times n} & 0 \\
         \end{array}
       \right)\,,~~~~\label{pfa-AA}
\eea
where
\bea
\W A\equiv \left(
         \begin{array}{c|c}
           ~~A_{(n+2)\times (n+2)}~~ &  A_{(n+2)\times n} \\
           \hline
          A_{n\times (n+2)} & 0 \\
         \end{array}
       \right)\,.~~~\label{WA}
\eea
When calculating \eref{pfa-AA}, a subtle point is that to avoid the ambiguity $\partial_{\epsilon_i\cdot k_+}$ acts on $\epsilon_i\cdot k_-$
and $\partial_{\epsilon_i\cdot k_-}$ acts on $\epsilon_i\cdot k_+$, one still need to rewrite $C_{ii}$ as in \eref{rewrite-C}.
After performing the operator ${\cal L}^\epsilon_\circ$, each $C_{ii}$ is turned to
\bea
C_{ii}\to\sum_{j\neq i,-}\,{(k_j\cdot k_i)z_{-j}\over z_{ji}z_{i-}}\,.
\eea
This object can be rewritten as
\bea
\sum_{j\in\{1,\cdots,n\}\setminus i}\,{k_i\cdot k_j\over z_{ij}}+{k_i\cdot \ell\over z_{i+}}-{k_i\cdot \ell\over z_{i-}}\,,
\eea
via the momentum conservation law and the massless condition $k_i^2=0$,
therefore vanishes automatically due to the $1$-loop scattering equations in \eref{SE-loop}.
Using the definition of Pfaffian \eref{pfa}, one can find that the non-vanishing contributions for \eref{pfa-AA} come from rows $i\in\{1,\cdots,n\}$ and columns $j\in\{n+1,\cdots,2n\}$, which give rise to the determinate of the matrix $A_{n\times n}$.
Thus the reduced Pfaffian of $\W A$
can be obtained as
\bea
{\bf Pf}'\W A=(-)^{{(n)(n-1)\over2}+1}{1\over z_{+-}}{\bf det}A_{n\times n}=(-)^{{n+2\over2}}z_{+-}({\bf Pf}'A)^2\,,~~~~\label{pf-WA}
\eea
where we have used $(-)^{n(n-1)\over2}=(-)^{n\over2}$, due to the fact $n$ is even. Combining \eref{eff-D-A} and \eref{pf-WA} together,
we arrive at
\bea
{\cal L}^\epsilon_\circ\,{\cal D}\,{\cal F}\,{\bf Pf}'\,\Psi=(-)^{{n+2\over2}}({\bf Pf}'A)^2\,,~~~~\label{eff-LD-A}
\eea
it shows that the operator ${\cal L}^\epsilon\,{\cal D}$ transmutes ${\cal F}\,{\bf Pf}'\,\Psi$ to $({\bf Pf}'A)^2$, up to an overall sign.
This result together with  the relations in \eref{rela-loop} indicate that the $1$-loop CHY integrands for BI, NLSM and SG are given as
\bea
& &{\cal I}^{L}_{\circ{\rm BI}}=({\bf Pf}'A)^2\,,~~~~~~~~{\cal I}^{R}_{\circ{\rm BI}}={\cal F}{\bf Pf}'\Psi\,,\nn
& &{\cal I}^{L}_{\circ{\rm NLSM}}=({\bf Pf}'A)^2\,,~~~~~~~~{\cal I}^{R}_{\circ{\rm NLSM}}=PT(\sigma_1\,\cdots,\sigma_n)\,,\nn
& &{\cal I}^{L}_{\circ{\rm SG}}=({\bf Pf}'A)^2\,,~~~~~~~~{\cal I}^{R}_{\circ{\rm SG}}=({\bf Pf}'A)^2\,.
\eea
The above integrands can also be obtained by using the forward limit method, this confirm that our argument about the operator
${\cal L}^\epsilon_\circ\,{\cal D}$ is correct.

Then, we consider applying the operator $\bar{{\cal L}}^\epsilon_\circ\,{\cal D}$ to ${\cal F}\,{\bf Pf}'\,\Psi$.
The first step also gives \eref{eff-D-A}. Then, we use the fact that each polarization vector appears once and only once in each term
of the reduced Pfaffian, it means the operator $\partial_{\epsilon_i\cdot\epsilon_j}$ turns $\epsilon_i\cdot\epsilon_j$ to $1$ and annihilates all other $\epsilon_i\cdot V$
and $\epsilon_j\cdot V$ simultaneously. This observation leads to
\bea
\bar{{\cal L}}^\epsilon_\circ\,{\bf Pf}'\,[\Psi_l]_{n,+,-:n}={\bf Pf}'\, \bar{A}={-1\over z_{+-}}\,{\bf Pf} \left(
         \begin{array}{c|c}
           ~~A_{n\times n}~~ &  0\\
           \hline
          0 & -A_{n\times n} \\
         \end{array}
       \right)\,,~~~~\label{pfa-A}
\eea
where
\bea
\bar{A}\equiv \left(
         \begin{array}{c|c}
           ~~A_{(n+2)\times (n+2)}~~ & 0 \\
           \hline
          0 & -A_{n\times n} \\
         \end{array}
       \right)\,.~~~\label{barA}
\eea
Evaluating \eref{eff-LD-A} directly, we also arrive at the result in \eref{eff-LD-A}. Thus we conclude that at the $1$-loop level
the operator $\bar{{\cal L}}^\epsilon_\circ$ is equivalent to ${\cal L}^\epsilon_\circ$ when applying to the object ${\cal F}\,{\bf Pf}'\,\Psi$,
similar as the situation at the tree level.

\subsection{EM and DBI}
\label{EM-DBI}

In this subsection, we consider the operator which transmutes the GR Feynman integrand to the EM Feynman integrand,
and transmutes the BI Feynman integrand to the DBI Feynman integrand.
Similar as the EYM and YMS case, the virtual particles running in the loop is not unique for the these cases, as can be seen
in Fig.\ref{virtual}. For the EYM and YMS cases, we have not treat all possibilities of the virtual loop particles, due to some
technic difficulties. But in this subsection, we will consider the full EM and DBI Feynman integrands with fixed external particles,
which include all possibilities of virtual particles.

We begin with the EM theory that photons carry no flavor. Before discussing the desired operator, let us take a look at the forward limit for the EM case. As pointed out in subsection \ref{forwardlimit},
the partial fraction identity indicates that each loop propagator should be cut once. The EM Feynman integrand is not color ordered, thus
it dose not require the cyclic summation of tree level color orderings. However, since
the virtual particles running in the loop can be either gravitons or photons, in order to cut each loop propagator once, one need to take the forward limit for each of two candidates once,
where one candidate is taking the forward limit for two external gravitons, and another one is for two external photons. Consequently, the $1$-loop EM Feynman integrand is obtained via
\bea
{\bf I}^{\epsilon,\W\epsilon}_{\rm EM}({\pmb{P}}_{2m};{\pmb H}_{n-2m})&=&{1\over \ell^2}\,\Big({\cal F}_g\,\,{\cal A}^{\epsilon,\W\epsilon}_{\rm EM}({\pmb P}_{2m};{\pmb H}_{n+2-2m})+{\cal F}_p\,{\cal A}^{\epsilon,\W\epsilon}_{\rm EM}({\pmb{P}}_{2m+2};{\pmb H}_{n-2m})\Big)\nn
&=&{1\over \ell^2}\,\Big({\cal F}_g\,{\cal T}^\epsilon_{X_{2m}}+{\cal F}_p\,{\cal T}^\epsilon_{X_{2m+2}}\Big)\,{\cal A}^{\epsilon,\W\epsilon}_{\rm GR}(\pmb{H}_{n+2})\,,~~~~\label{I-EM}
\eea
where ${\cal F}_g$ and ${\cal F}_p$ mean taking forward limit for two gravitons and two photons, respectively.
The definition of the operator ${\cal T}^\epsilon_{X_{2m}}$ can be seen in \eref{TX2}, and we have used the relation in Table.\ref{tab:unifying}
that ${\cal T}^\epsilon_{X_{2m}}$ transmutes the tree GR amplitude to the EM one.
Without lose of generality, let us assume
the operator ${\cal T}^\epsilon_{X_{2m}}$ is defined for the length-$2m$ set $\{1,\cdots,2m\}$, while the operator ${\cal T}^\epsilon_{X_{2m+2}}$ is defined for $\{1,\cdots,2m,+,-\}$.

Now we argue that the operator ${\cal T}^\epsilon_{X_{2m}}({\cal D}+1)$ transmutes the $1$-loop GR Feynman integrand ${\bf I}^{\epsilon,\W\epsilon}_{\rm GR}({\pmb H}_n)=(1/\ell^2){\cal F}\,{\cal A}^{\epsilon,\W\epsilon}_{\rm GR}({\pmb H}_{n+2})$ to ${\bf I}^{\epsilon,\W\epsilon}_{\rm EM}({\pmb P}_{2m};{\pmb H}_{n-2m})$.  As will be seen soon, applying ${\cal T}^\epsilon_{X_{2m}}{\cal D}$ to ${\bf I}^{\epsilon,\W\epsilon}_{\rm GR}({\pmb H}_n)$ gives a part of ${\cal F}_p\,{\cal T}^\epsilon_{X_{2m+2}}\,{\cal A}^{\epsilon,\W\epsilon}_{\rm GR}({\pmb H}_{n+2})$ in \eref{I-EM}, while applying ${\cal T}^\epsilon_{X_{2m}}$ to ${\bf I}^{\epsilon,\W\epsilon}_{\rm GR}({\pmb H}_n)$ gives ${\cal F}_g\,{\cal T}^\epsilon_{X_{2m}}\,{\cal A}^{\epsilon,\W\epsilon}_{\rm GR}({\pmb H}_{n+2})$ and the remaining part of ${\cal F}_p\,{\cal T}^\epsilon_{X_{2m+2}}\,{\cal A}^{\epsilon,\W\epsilon}_{\rm GR}({\pmb H}_{n+2})$.

To show this, we first decompose the operator ${\cal T}^\epsilon_{X_{2m+2}}$ into two pieces as follows
\bea
{\cal T}^\epsilon_{X_{2m+2}}={\cal T}^\epsilon_{X_{2m}}\,{\cal T}^\epsilon[\overline{+,-}]
+\sum_{\substack{i,j\in\{1,\cdots,n\}\\i\neq j}}\,{\cal T}^\epsilon_{X_{2m-2}^{ij}}\,{\cal T}^\epsilon[\overline{i,+}]\,{\cal T}^\epsilon[\overline{j,-}]\,,~~~~\label{separa}
\eea
then the object ${\cal F}_p\,{\cal T}^\epsilon_{X_{2m+2}}\,{\cal A}^{\epsilon,\W\epsilon}_{\rm GR}({\pmb H}_{n+2})$ in \eref{I-EM}
is also separated into two parts. Here the operator ${\cal T}^\epsilon_{X_{2m-2}^{ij}}$ is defined for the length-$(2m-2)$ set $\{1,\cdots,2m\}\setminus\{i,j\}$.
For the first part, using the relation \eref{commu-D-F}, we obtain
\bea
{\cal T}^\epsilon_{X_{2m}}\,{\cal D}\,{\cal F}\,{\cal A}^{\epsilon,\W\epsilon}_{\rm GR}({\pmb H}_{n+2})={\cal F}_p\,{\cal T}^\epsilon_{X_{2m}}\,{\cal T}^\epsilon[\overline{+,-}]\,{\cal A}^{\epsilon,\W\epsilon}_{\rm GR}({\pmb H}_{n+2})\,,~~~~\label{TD-EM}
\eea
where the commutativity between ${\cal T}^\epsilon_{X_{2m}}$ and ${\cal F}_p$ have been used. Notice that the virtual particle running in the loop
is fixed to be a photon after performing the operator ${\cal D}$, similar as in the EYM case discussed in subsection \ref{ssEYM-ssYMS}.

Now we turn to the second part.
To treat this part, we observe that at the tree level the operator ${\cal T}^\epsilon[\overline{i,+}]\,{\cal T}^\epsilon[\overline{j,-}]$ turns $(\epsilon_i\cdot\epsilon_+)(\epsilon_j\cdot \epsilon_-)$ to $1$, and annihilates all other terms do not contain $(\epsilon_i\cdot\epsilon_+)(\epsilon_j\cdot \epsilon_-)$. At the $1$-loop level, the corresponding object behaves as
\bea
\sum_{r}\,(\epsilon_i\cdot\epsilon^r_+)(\epsilon^r_-\cdot \epsilon_j)=\epsilon_i\cdot\epsilon_j\,,~~~~\label{eiej}
\eea
due to the forward limit procedure.
Notice that in general the summation $\sum_r(\epsilon_+^{r})^\mu(\epsilon_-^{r})^\nu$ should be
\bea
\sum_r\,(\epsilon_+^{r})^\mu(\epsilon_-^{r})^\nu=\eta^{\mu\nu}-{l^\mu q^\nu+l^\nu q^\mu\over l\cdot q}\equiv\Delta^{\mu\nu}\,,
\eea
where the null $q$ satisfies $\epsilon_+\cdot q=\epsilon_-\cdot q=0$.
Here we are allowed to drop the $q$-dependent term in $\Delta^{\mu\nu}$, since its contribution vanishes on the solutions
to the scattering equations; see \cite{Roehrig:2017gbt}.
Thus, the effect of applying ${\cal T}^\epsilon[\overline{i,}+]\,{\cal T}^\epsilon[\overline{j,-}]$ at tree level
is associated to applying ${\cal T}^\epsilon[\overline{i,j}]$ at $1$-loop level. But one can not conclude that
\bea
{\cal F}_p\,{\cal T}^\epsilon[\overline{i,+}]\,{\cal T}^\epsilon[\overline{j,-}]\,{\cal A}^{\epsilon,\W\epsilon}_{\rm GR}({\pmb H}_{n+2})&=&{\cal T}^\epsilon[\overline{i,j}]\,{\cal F}\,{\cal A}^{\epsilon,\W\epsilon}_{\rm GR}({\pmb H}_{n+2})\,.
\eea
The first reason is, the object $\epsilon_i\cdot\epsilon_j$ at $1$-loop level receives contributions from either $(\epsilon_i\cdot\epsilon_+)(\epsilon_j\cdot \epsilon_-)$ or $(\epsilon_i\cdot\epsilon_-)(\epsilon_j\cdot \epsilon_+)$ at tree level,
thus turning $\epsilon_i\cdot\epsilon_j$ to $1$ at $1$-loop level corresponds to turning both $(\epsilon_i\cdot\epsilon_+)(\epsilon_j\cdot \epsilon_-)$ and $(\epsilon_i\cdot\epsilon_-)(\epsilon_j\cdot \epsilon_+)$ to $1$ at tree level.
The second reason is, the Lorentz invariant $(\epsilon_i\cdot \epsilon_j)$
in ${\bf I}^{\epsilon,\W\epsilon}_{\rm GR}({\pmb H}_{n})$ has two origins. Except $(\epsilon_i\cdot\epsilon_+)(\epsilon_j\cdot \epsilon_-)$ or $(\epsilon_i\cdot\epsilon_-)(\epsilon_j\cdot \epsilon_+)$ in the tree amplitude,  $(\epsilon_i\cdot \epsilon_j)$ in the tree amplitude ${\cal A}^{\epsilon,\W\epsilon}_{\rm GR}({\pmb H}_{n+2})$ also causes $(\epsilon_i\cdot \epsilon_j)$ in ${\bf I}^{\epsilon,\W\epsilon}_{\rm GR}({\pmb H}_{n})$. The operator $\partial_{\epsilon_i\cdot\epsilon_j}$ can not distinguish these two origins.
Motivated by the above discussion, we separate ${\cal A}^{\epsilon,\W\epsilon}_{\rm GR}({\pmb H}_{n+2})$ into three parts ${\cal A}^{\epsilon,\W\epsilon}_{\rm GR;1}({\pmb H}_{n+2})$,
${\cal A}^{\epsilon,\W\epsilon}_{\rm GR;2}({\pmb H}_{n+2})$ and ${\cal A}^{\epsilon,\W\epsilon}_{\rm GR;3}({\pmb H}_{n+2})$, where ${\cal A}^{\epsilon,\W\epsilon}_{\rm GR;1}({\pmb H}_{n+2})$ contains $(\epsilon_i\cdot\epsilon_+)(\epsilon_j\cdot \epsilon_-)$ or $(\epsilon_i\cdot\epsilon_-)(\epsilon_j\cdot \epsilon_+)$, ${\cal A}^{\epsilon,\W\epsilon}_{\rm GR;2}({\pmb H}_{n+2})$ contains $(\epsilon_i\cdot\epsilon_j)$, while ${\cal A}^{\epsilon,\W\epsilon}_{\rm GR;3}({\pmb H}_{n+2})$ contains none of $(\epsilon_i\cdot\epsilon_+)(\epsilon_j\cdot \epsilon_-)$, $(\epsilon_i\cdot\epsilon_-)(\epsilon_j\cdot \epsilon_+)$ and $(\epsilon_i\cdot\epsilon_j)$.
Since each polarization vector appears once and only once in each term of the amplitude, the above three parts have no overlap. Then we have
\bea
& &{\cal F}_p\,\Big({\cal T}^\epsilon[\overline{i,+}]\,{\cal T}^\epsilon[\overline{j,-}]+{\cal T}^\epsilon[\overline{j,+}]\,{\cal T}^\epsilon[\overline{i,-}]\Big)\,{\cal A}^{\epsilon,\W\epsilon}_{\rm GR;1}({\pmb H}_{n+2})={\cal T}^\epsilon[\overline{i,j}]\,{\cal F}\,{\cal A}^{\epsilon,\W\epsilon}_{\rm GR;1}({\pmb H}_{n+2})\,,\nn
& &{\cal T}^\epsilon[\overline{i,+}]\,{\cal T}^\epsilon[\overline{j,-}]\,{\cal A}^{\epsilon,\W\epsilon}_{\rm GR;2}({\pmb H}_{n+2})={\cal T}^\epsilon[\overline{i,+}]\,{\cal T}^\epsilon[\overline{j,-}]\,{\cal A}^{\epsilon,\W\epsilon}_{\rm GR;3}({\pmb H}_{n+2})=0\,.~~~~\label{T-EM1}
\eea
On the other hand, we have
\bea
& &{\cal F}_g\,{\cal T}^\epsilon[\overline{i,j}]\,{\cal A}^{\epsilon,\W\epsilon}_{\rm GR;2}({\pmb H}_{n+2})={\cal T}^\epsilon[\overline{i,j}]\,{\cal F}\,{\cal A}^{\epsilon,\W\epsilon}_{\rm GR;2}({\pmb H}_{n+2})\,,\nn
& &{\cal T}^\epsilon[\overline{i,j}]\,{\cal A}^{\epsilon,\W\epsilon}_{\rm GR;1}({\pmb H}_{n+2})={\cal T}^\epsilon[\overline{i,j}]\,{\cal A}^{\epsilon,\W\epsilon}_{\rm GR;3}({\pmb H}_{n+2})=0\,.~~~~\label{T-EM2}
\eea
Combining \eref{T-EM1} and \eref{T-EM2} together gives
\bea
{\cal T}^\epsilon[\overline{i,j}]\,{\cal F}\,{\cal A}^{\epsilon,\W\epsilon}_{\rm GR}({\pmb H}_{n+2})&=&{\cal F}_p\Big({\cal T}^\epsilon[\overline{i,+}]\,{\cal T}^\epsilon[\overline{j,-}]+{\cal T}^\epsilon[\overline{j,+}]\,{\cal T}^\epsilon[\overline{i,-}]\Big)\,{\cal A}^{\epsilon,\W\epsilon}_{\rm GR;1}({\pmb H}_{n+2})+{\cal F}_g\,{\cal T}^\epsilon[\overline{i,j}]\,{\cal A}^{\epsilon,\W\epsilon}_{\rm GR;2}({\pmb H}_{n+2})\nn
&=&\Big\{{\cal F}_p\Big({\cal T}^\epsilon[\overline{i,+}]\,{\cal T}^\epsilon[\overline{j,}-]+{\cal T}^\epsilon[\overline{j,}+]\,{\cal T}^\epsilon[\overline{i,-}]\Big)+{\cal F}_g\,{\cal T}^\epsilon[\overline{i,j}]\Big\}\,{\cal A}^{\epsilon,\W\epsilon}_{\rm GR}({\pmb H}_{n+2})\,,~~~~\label{T-EM}
\eea
where we have used the observation
\bea
{\cal T}^\epsilon[\overline{i,j}]\,{\cal F}\,{\cal A}^{\epsilon,\W\epsilon}_{\rm GR;3}({\pmb H}_{n+2})=0\,,
\eea
arise from the fact that ${\cal F}\,{\cal A}^{\epsilon,\W\epsilon}_{\rm GR;3}({\pmb H}_{n+2})$ does not contain $(\epsilon_i\cdot\epsilon_j)$.
Using \eref{T-EM}, we have
\bea
& &\sum_{\substack{i,j\in\{1,\cdots,n\}\\i< j}}\,{\cal T}^\epsilon_{X_{2m-2}^{ij}}\,{\cal T}^\epsilon[\overline{i,j}]\,{\cal F}\,{\cal A}^{\epsilon,\W\epsilon}_{\rm GR}({\pmb H}_{n+2})\nn
&=&\sum_{\substack{i,j\in\{1,\cdots,n\}\\i< j}}\,{\cal F}_p\,{\cal T}^\epsilon_{X_{2m-2}^{ij}}\,\Big({\cal T}^\epsilon[\overline{i,+}]\,{\cal T}^\epsilon[\overline{j,-}]+{\cal T}^\epsilon[\overline{j,+}]\,{\cal T}^\epsilon[\overline{i,}-]\Big)\,{\cal A}^{\epsilon,\W\epsilon}_{\rm GR}({\pmb H}_{n+2})\nn
& &+\sum_{\substack{i,j\in\{1,\cdots,n\}\\i< j}}\,{\cal F}_g\,{\cal T}^\epsilon_{X_{2m-2}^{ij}}\,{\cal T}^\epsilon[\overline{i,j}]\,{\cal A}^{\epsilon,\W\epsilon}_{\rm GR}({\pmb H}_{n+2})\,,~~~~\label{T-EM-f}
\eea
where the manifest commutativity between ${\cal T}^\epsilon_{X_{2m-2}^{ij}}$ and ${\cal F}_p$, ${\cal F}_g$ has been used.
It is straightforward to recognize that
\bea
& &\sum_{\substack{i,j\in\{1,\cdots,n\}\\i< j}}\,{\cal T}^\epsilon_{X_{2m-2}^{ij}}\,{\cal T}^\epsilon[\overline{i,j}]={\cal T}^\epsilon_{X_{2m}}\,,\nn
& &\sum_{\substack{i,j\in\{1,\cdots,n\}\\i< j}}\,{\cal T}^\epsilon_{X_{2m-2}^{ij}}\,\Big({\cal T}^\epsilon[\overline{i,+}]\,{\cal T}^\epsilon[\overline{j,-}]+{\cal T}^\epsilon[\overline{j,+}]\,{\cal T}^\epsilon[\overline{i,-}]\Big)
=\sum_{\substack{i,j\in\{1,\cdots,n\}\\i\neq j}}\,{\cal T}^\epsilon_{X_{2m-2}^{ij}}\,{\cal T}^\epsilon[\overline{i,+}]\,{\cal T}^\epsilon[\overline{j,-}]\,.
\eea
Substituting this into  \eref{T-EM-f}, we immediately get
\bea
{\cal T}^\epsilon_{X_{2m}}\,{\cal F}\,{\cal A}^{\epsilon,\W\epsilon}_{\rm GR}({\pmb H}_{n+2})
&=&{\cal F}_p\,\Big(\sum_{\substack{i,j\in\{1,\cdots,n\}\\i\neq j}}\,{\cal T}^\epsilon_{X_{2m-2}^{ij}}\,{\cal T}^\epsilon[\overline{i,+}]\,{\cal T}^\epsilon[\overline{j,-}]\Big)\,{\cal A}^{\epsilon,\W\epsilon}_{\rm GR}({\pmb H}_{n+2})+{\cal F}_g\,{\cal T}^\epsilon_{X_{2m}}\,{\cal A}^{\epsilon,\W\epsilon}_{\rm GR}({\pmb H}_{n+2})\,.~~~~\label{T-EM-ff}
\eea
The operator in the big bracket at the r.h.s is just the remaining operator in \eref{separa}, this part together with \eref{TD-EM} gives the
second term in \eref{I-EM}. The second part at the r.h.s of \eref{T-EM-ff} gives the first term in \eref{I-EM}.
Putting these together, we finally find
\bea
{\cal T}^\epsilon_{X_{2m}}({\cal D}+1)\,{\cal F}\,{\cal A}^{\epsilon,\W\epsilon}_{\rm GR}({\pmb H}_{n+2})=\Big({\cal F}_g{\cal T}^\epsilon_{X_{2m}}+{\cal F}_P\,{\cal T}^\epsilon_{X_{2m+2}}\Big)\,{\cal A}^{\epsilon,\W\epsilon}_{\rm GR}({\pmb H}_{n+2})\,,~~~\label{generalization-OF-FO}
\eea
and
\bea
{\cal T}^\epsilon_{X_{2m}}({\cal D}+1)\,{\bf I}^{\epsilon,\W\epsilon}_{\rm GR}({\pmb H}_{n})={\bf I}^{\epsilon,\W\epsilon}_{\rm EM}({\pmb P}_{2m};{\pmb H}_{n-2m})\,.~~~~\label{OP-EM}
\eea
The relation \eref{generalization-OF-FO} serves as the generalization of our basic idea ${\cal O}_\circ\,{\cal F}\,\bullet={\cal F}\,{\cal O}\,\bullet$.

The above relation can be verified by applying the operator ${\cal T}^\epsilon_{X_{2m}}({\cal D}+1)$ to ${\bf I}^{\epsilon,\W\epsilon}_{\rm GR}({\pmb H}_{n})$,
and comparing the resulting object to ${\bf I}^{\epsilon,\W\epsilon}_{\rm EM}({\pmb P}_{2m};{\pmb H}_{n-2m})$ obtained through the forward limit method, similar as what we did in subsection \ref{BI,NLSM,SG}.

Now we move to the EMf theory that photons carry flavors. For this case, another new situation occurs. As pointed out in subsection \ref{forwardlimit}, when taking the forward limit, one should sum over the allowed flavors of legs $+$ and $-$ if they are photons.
Thus the relation \eref{I-EM} should be modified as
\bea
{\bf I}^{\epsilon,\W\epsilon}_{\rm EMf}({\pmb{P}}_{2m};{\pmb H}_{n-2m})={1\over \ell^2}\,\Big({\cal F}_g\,{\cal T}^\epsilon_{{\cal X}_{2m}}+\sum_{I_+,I_-}\,\delta_{I_+I_-}\,{\cal F}_p\,{\cal T}^\epsilon_{{\cal X}_{2m+2}}\Big)\,{\cal A}^{\epsilon,\W\epsilon}_{\rm GR}(\pmb{H}_{n+2})\,,~~~~\label{I-EM-2}
\eea
where we have used the condition that when summing over the flavors of two internal particles, two flavors must be identified. Similar as in \eref{separa},
we separate the operator ${\cal T}^\epsilon_{{\cal X}_{2m+2}}$ as
\bea
{\cal T}^\epsilon_{{\cal X}_{2m+2}}={\cal T}^\epsilon_{{\cal X}_{2m}}\,\delta_{I_+I_-}\,{\cal T}^\epsilon[\overline{+,-}]
+\sum_{\substack{i,j\in\{1,\cdots,n\}\\i\neq j}}\,{\cal T}^\epsilon_{{\cal X}_{2m-2}^{ij}}\,\delta_{I_iI_+}\,\delta_{I_jI_-}\,{\cal T}^\epsilon[\overline{i,+}]\,{\cal T}^\epsilon[\overline{j,-}]\,.~~~\label{separa-2}
\eea
Substituting the separation \eref{separa-2} into \eref{I-EM-2}, and doing the summation over $I_+$ and $I_-$, we get
\bea
& &{\bf I}^{\epsilon,\W\epsilon}_{\rm EMf}({\pmb{P}}_{2m};{\pmb H}_{n-2m})\nn
&=&{1\over \ell^2}\,\Big({\cal F}_g\,{\cal T}^\epsilon_{{\cal X}_{2m}}+N\,{\cal F}_p\,{\cal T}^\epsilon_{{\cal X}_{2m}}\,{\cal T}^\epsilon[\overline{+,-}]+{\cal F}_p\,\sum_{\substack{i,j\in\{1,\cdots,n\}\\i\neq j}}\,{\cal T}^\epsilon_{{\cal X}_{2m-2}^{ij}}\,\delta_{I_iI_j}\,{\cal T}^\epsilon[\overline{i,+}]\,{\cal T}^\epsilon[\overline{j,-}]\Big)\,{\cal A}^{\epsilon,\W\epsilon}_{\rm GR}(\pmb{H}_{n+2})\,,
\eea
where $N$ stands for the number of different flavors.
Using the previous technics, we can recognize these terms as
\bea
{1\over \ell^2}\,\Big(N\,{\cal F}_p\,{\cal T}^\epsilon_{{\cal X}_{2m}}\,{\cal T}^\epsilon[\overline{+,-}]\Big)\,{\cal A}^{\epsilon,\W\epsilon}_{\rm GR}(\pmb{H}_{n+2})={1\over \ell^2}\,\Big(N\,{\cal T}^\epsilon_{{\cal X}_{2m}}\,{\cal D}\Big)\,{\cal F}\,{\cal A}^{\epsilon,\W\epsilon}_{\rm GR}(\pmb{H}_{n+2})\,,
\eea
and
\bea
{1\over \ell^2}\,\Big({\cal F}_g\,{\cal T}^\epsilon_{{\cal X}_{2m}}+{\cal F}_p\,\sum_{\substack{i,j\in\{1,\cdots,n\}\\i\neq j}}\,{\cal T}^\epsilon_{{\cal X}_{2m-2}^{ij}}\,\delta_{I_iI_j}\,{\cal T}^\epsilon[\overline{i,+}]\,{\cal T}^\epsilon[\overline{j,-}]\Big)\,{\cal A}^{\epsilon,\W\epsilon}_{\rm GR}(\pmb{H}_{n+2})
={1\over \ell^2}\,{\cal T}^\epsilon_{{\cal X}_{2m}}\,{\cal F}\,{\cal A}^{\epsilon,\W\epsilon}_{\rm GR}(\pmb{H}_{n+2})\,.
\eea
Combining them together, we find that the operator ${\cal T}^\epsilon_{{\cal X}_{2m}}(N{\cal D}+1)$ transmutes the $1$-loop GR Feynman integrand to the $1$-loop EMf Feynman integrand, formulated as
\bea
{\cal T}^\epsilon_{{\cal X}_{2m}}(N{\cal D}+1)\,{\bf I}^{\epsilon,\W\epsilon}_{\rm GR}({\pmb H}_{n})={\bf I}^{\epsilon,\W\epsilon}_{\rm EMf}({\pmb P}_{2m};{\pmb H}_{n-2m})\,.~~~~\label{OP-EMf}
\eea

The underlying basic of relation \eref{OP-EMf} is
\bea
{\cal T}^\epsilon_{{\cal X}_{2m}}(N{\cal D}+1)\,{\cal F}\,{\bf Pf}'\Psi=\Big({\cal F}_g{\cal T}^\epsilon_{{\cal X}_{2m}}+\sum_{I_+,I_-}\,\delta_{I_+I_-}\,{\cal F}_P\,{\cal T}^\epsilon_{{\cal X}_{2m+2}}\Big)\,{\bf Pf}'\Psi\,,
\eea
which can also be applied to the BI and DBI theory, due to the tree level relations in Table.\ref{tab:unifying}. Thus we also have the relation
\bea
{\cal T}^\epsilon_{{\cal X}_{2m}}(N{\cal D}+1)\,{\bf I}^{\epsilon}_{\rm BI}({\pmb P}_{n})={\bf I}^{\epsilon}_{\rm DBI}({\pmb S}_{2m};{\pmb P}_{n-2m})\,.~~~~\label{OP-DBI}
\eea
%

\section{Factorization of operators}
\label{sectionfac-op}

In this section, we discuss the factorization of differential operators constructed in the previous two sections.
Under the unitarity cut, the $1$-loop Feynman integrand factorizes into two on-shell tree amplitudes, and we will show that the associated
$1$-loop level operator factorizes into two tree level operators parallelly. As a by product, we also demonstrate that the tree level operator factorizes into tree level operators with lower points, paralleled to the factorization of tree amplitudes. The above statement implies that
both the $1$-loop and tree operators can be verified by applying the tree level operator to the tree amplitude with lowest points. The structure of this section is as follows. In subsection \ref{general}, we give the general discussion about the unitarity cut and the factorizations of operators. Then, we show the factorizations of three operators ${\cal T}^\epsilon_{\circ C}[\overline{\sigma_1,\cdots,\sigma_n}]$, ${\cal L}^\epsilon_\circ\,{\cal D}$ and ${\cal T}^\epsilon_{X_m}\,({\cal D}+1)$ in subsections
\ref{OP-1}, \ref{OP-2} and \ref{OP-3}, respectively.

\subsection{General discussion}
\label{general}

As well known, the $1$-loop Feynman integrand factorizes into two tree amplitudes under the so called unitarity cut operation.
For the $s_P$ channel with respecting to the cut momentum $P$, the unitarity cut is evaluated as
\bea
\Delta{\cal A}_\circ|_P=\int\,d\Omega\,\W{\bf I}|_P\,,
\eea
where $\W{\bf I}|_P\equiv \ell^2(\ell+P)^2\,{\bf I}$ is the cut integrand obtained via multiplying the full $1$-loop Feynman integrand
by two cut propagators $\ell^2$ and $(\ell+P)^2$. Here we use ${\cal A}_\circ$ to denote the $1$-loop amplitude. The measure above is given by
\bea
\int\,d\Omega &=&\int\,d^d\ell\,\delta(\ell^2)\delta((\ell+P)^2)=\int\,d^d\ell\,d^d\W \ell\,\delta(\ell^2)\delta(\W \ell^2)\delta^d(\W \ell-\ell-P)\,.
\eea
Under the constraints of $\delta$-functions, $\W{\bf I}|_P$ is factorized as
\bea
\lim_{\substack{\ell^2\to 0\\(\ell+P)^2\to 0}}\,\W{\bf I}|_P=\sum_{h_1,h_2}\,{\cal A}_L(\cdots,(\ell+P)^{h_1},-\ell^{h_2})\,{\cal A}_R(\cdots,\ell^{\bar{h}_2},-(\ell+P)^{\bar{h}_2}) \,.
\eea
Here $\sum_{h_1,h_2}$ means sum over all possible polarization vectors or polarization tensors carried by the  external legs arise from cutting the internal loop propagator.
Suppose the $1$-loop Feynman integrand ${\bf I}$ for the considered theory can be generated from the $1$-loop Feynman integrand of another theory via the
operator ${\cal O}_\circ$, i.e., ${\bf I}={\cal O}_\circ\,{\bf I}'$, and the tree amplitude can be generated via the operator ${\cal O}$, then, under the constraints
of $\delta$-functions, we have
\bea
\lim_{\substack{\ell^2\to 0\\(\ell+P)^2\to 0}}\,\ell^2(\ell+P)^2\,{\cal O}_\circ\,{\bf I}'=\sum_{h_1,h_2}\,\Big({\cal O}_L\,{\cal A}'_L(\cdots,(\ell+P)^{h'_2},-\ell^{h'_1})\Big)\,\Big({\cal O}_R\,{\cal A}'_R(\cdots,\ell^{\bar{h}'_1},-(\ell+P)^{\bar{h}'_2})\Big) \,.~~~~\label{fac-op}
\eea
In other words, if we multiply the resulting Feynman integrand by two
corresponding propagators, and setting two virtual particles running in the loop to be on-shell, the $1$-loop level operator ${\cal O}_\circ$ will factorize into two tree level operators ${\cal O}_L$ and ${\cal O}_R$,
where the operator ${\cal O}_L$ acts on ${\cal A}'_L$ and annihilates ${\cal A}'_R$, while the operator ${\cal O}_R$ acts on ${\cal A}'_R$ and annihilates ${\cal A}'_L$. We emphasize that this factorization of the operator does not mean the operator ${\cal O}_\circ$ can be decomposed into ${\cal O}_L$ and ${\cal O}_R$ algebraically. It only means the l.h.s and r.h.s of \eref{fac-op} are equivalent.
It is interesting to take a look at how such factorization realize.

To understand the factorization of the operator, let us separate the operation of unitarity cut into two steps
\bea
\lim_{\substack{\ell^2\to 0\\(\ell+P)^2\to 0}}\,\ell^2(\ell+P)^2\,{\bf I}=\lim_{(\ell+P)^2\to 0}\,(\ell+P)^2\,\Big(\lim_{\ell^2\to 0}\,\ell^2{\bf I}\Big)\,.~~~~\label{2-steps}
\eea
Without lose of generality, we can identify the loop momentum $\ell$ in the above formula as $\ell$ in the forward limit method. Then, the forward limit method indicates that
\bea
\lim_{\ell^2\to 0}\,\ell^2{\bf I}=\sum_{h_1}\,{\cal A}(\cdots,-\ell^{h_1},\ell^{\bar{h}_1})\,,~~~~\label{reverse-forward}
\eea
where ${\cal A}(\cdots,-\ell^{h_1},\ell^{\bar{h}_1})$ is the tree amplitude includes two on-shell external momenta $-\ell$ and $\ell$.
Now we assume the Feynman integrand ${\bf I}$ can be generated from ${\bf I}'$ via ${\bf I}={\cal O}_\circ\,{\bf I}'$. For theories under consideration in this paper, our results show that \eref{reverse-forward} can be written as
\bea
\lim_{\ell^2\to 0}\,\ell^2{\cal O}_\circ\,{\bf I}'=\sum_{h_1}\,{\cal O}\,{\cal A}'(\cdots,-\ell^{h'_1},\ell^{\bar{h}'_1})\,.~~~~\label{rewitten-reverse-forward}
\eea
Substituting this into \eref{2-steps}, then substituting \eref{2-steps} into \eref{fac-op}, we find that the remaining work is to show that
the tree level operator ${\cal O}$ factorizes into ${\cal O}_L$ and ${\cal O}_R$ as
\bea
& &\lim_{(\ell+P)^2\to 0}\,(\ell+P)^2\,{\cal O}\,{\cal A}'(\cdots,-\ell^{h'_1},\ell^{\bar{h}'_1})\nn
&=&\sum_{h_2}\,\Big({\cal O}_L\,{\cal A}'_L(\cdots,(\ell+P)^{h'_2},-\ell^{h'_1})\Big)\,\Big({\cal O}_R\,{\cal A}'_R(\cdots,\ell^{\bar{h}'_1},-(\ell+P)^{\bar{h}'_2})\Big) \,.
\eea
This is nothing but the correct factorization of the tree amplitude, i.e.,
\bea
\lim_{(\ell+P)^2\to 0}\,(\ell+P)^2\,{\cal A}(\cdots,-\ell^{h_1},\ell^{\bar{h}_1})
&=&\sum_{h_2}\,{\cal A}_L(\cdots,(\ell+P)^{h_2},-\ell^{h_1})\,{\cal A}_R(\cdots,\ell^{\bar{h}_1},-(\ell+P)^{\bar{h}_2})\,,
\eea
thus must be correct.

\subsection{Operator ${\cal T}^\epsilon_{\circ C}[\overline{\sigma_1,\cdots,\sigma_n}]$}
\label{OP-1}

To illustrate the idea more clear, let us take the operator ${\cal T}^\epsilon_{\circ C}[\overline{\sigma_1,\cdots,\sigma_n}]$
as the example. For our purpose, it is sufficient to consider ${\bf I}^{\W\epsilon}_{\rm YM}={\cal T}^\epsilon_{\circ C}[\overline{\sigma_1,\cdots,\sigma_n}]\,{\bf I}^{\epsilon,\W\epsilon}_{\rm GR}$. Then \eref{rewitten-reverse-forward} becomes
\bea
\lim_{\ell^2\to 0}\,\ell^2{\cal T}^\epsilon_{\circ C}[\overline{\sigma_1,\cdots,\sigma_n}]\,{\bf I}^{\epsilon,\W\epsilon}_{\rm GR}({\pmb H}_n)&=&\sum_{h_1}\,\sum_i\,{\cal T}^\epsilon[\overline{+,\sigma_i,\cdots,\sigma_{i-1},-}]\,{\cal A}^{\epsilon,\W\epsilon}_{\rm GR}({\pmb H}_n+\{-\ell^{h'_1},\ell^{\bar{h}'_1}\})\nn
&=&\sum_{h_1}\,\sum_i\,{\cal A}^{\W\epsilon}_{\rm YM}\big(\overline{(+)^{\bar{h}_1},\sigma_i,\cdots,\sigma_{i-1},(-)^{h_1}}\big)\,.
\eea
Now we do the second step in \eref{2-steps} for the resulting object above,
\bea
\lim_{(\ell+P)^2\to 0}\,(\ell+P)^2\,\Big(\sum_{h_1}\,\sum_i\,{\cal A}^{\W\epsilon}_{\rm YM}\big(\overline{(+)^{\bar{h}_1},\sigma_i,\cdots,\sigma_{i-1},(-)^{h_1}}\big)\Big)\,.
\eea
For the color ordered amplitude, we can assume $P=\sum_{a=k}^l\,k_{\sigma_a}$. It is straightforward to observe that only the $i=k$
term in the summation over $i$ provides non-vanishing contribution, due to the constraint $(\ell+P)^2=0$. Thus our current aim is to show the relation
\bea
& &\lim_{(\ell+P)^2\to 0}\,(\ell+P)^2\,{\cal T}^\epsilon[\overline{+,\sigma_k,\cdots,\sigma_{k-1},-}]\,{\cal A}^{\epsilon,\W\epsilon}_{\rm GR}({\pmb H}_n+\{-\ell^{h'_1},\ell^{\bar{h}'_1}\})\nn
&=&\sum_{h_2}\,\Big({\cal O}_L\,{\cal A}^{\epsilon,\W\epsilon}_{{\rm GR};L}(k_{\sigma_l},\cdots,k_{\sigma_{k-1}},(\ell+P)^{h'_2},-\ell^{h'_1})\Big)\,\Big({\cal O}_R\,{\cal A}^{\epsilon,\W\epsilon}_{{\rm GR};R}(k_{\sigma_k},\cdots,k_{\sigma_{l-1}},-(\ell+P)^{\bar{h}'_2},\ell^{\bar{h}'_1})\Big)\,.
\eea
This is just the tree level factorization
\bea
& &\lim_{(\ell+P)^2\to 0}\,(\ell+P)^2\,{\cal A}^{\W\epsilon}_{\rm YM}\big(\overline{(+)^{\bar{h}_1},\sigma_k,\cdots,\sigma_{k-1},(-)^{h_1}}\big)\nn
&=&\sum_{h_2}\,{\cal A}^{\W\epsilon}_{{\rm YM};L}\big(\overline{(P_+)^{h_2},\sigma_l,\cdots,\sigma_{k-1},(-)^{h_1}}\big)\,{\cal A}^{\W\epsilon}_{{\rm YM};R}\big(\overline{(+)^{\bar{h}_1},\sigma_k,\cdots,\sigma_{l-1},(P_-)^{\bar{h}_2}}\big)\,,
\eea
thus must be true. Here the legs with momenta $(\ell+P)$ and $-(\ell+P)$ are denoted by $P_+$ and $P_-$ in the color orderings, respectively.

Although the factorization of the operator ${\cal T}^\epsilon[\overline{+,\sigma_k,\cdots,\sigma_{k-1},-}]$ at tree level is ensured by the factorization of tree amplitude, there is no harm to see how it realize. The following treatment is similar as that in \cite{Zhou:2020llv}. Let us choose the formula of operator ${\cal T}^\epsilon[\overline{+,\sigma_k,\cdots,\sigma_{k-1},-}]$ to be
\bea
{\cal T}^\epsilon[\overline{+,\sigma_k,\cdots,\sigma_{k-1},-}]=\Big(\prod_{i=k}^{k-2}\,{\cal I}^\epsilon_{+\sigma_i\sigma_{i+1}}\Big)\,{\cal I}^\epsilon_{+\sigma_{k-1}-}\,{\cal T}^\epsilon[\overline{+,-}]\,,
\eea
and denote the polarization vectors associated to $(\ell+P)$ and $-(\ell+P)$ by $\epsilon_{P_+}$ and $\epsilon_{P_-}$, respectively.
We apply the operator ${\cal T}^\epsilon[\overline{+,\sigma_k,\cdots,\sigma_{k-1},-}]$ chosen above to the factorized formula of ${\cal A}^{\epsilon,\W\epsilon}_{\rm GR}({\pmb H}_n+\{-\ell^{h'_1},\ell^{\bar{h}'_1}\})$ which is given as
\bea
& &\lim_{(\ell+P)^2\to 0}\,(\ell+P)^2\,{\cal A}^{\epsilon,\W\epsilon}_{\rm GR}({\pmb H}_n+\{-\ell^{h'_1},\ell^{\bar{h}'_1}\})\nn
&=&\sum_{h'_2}\,{\cal A}^{\epsilon,\W\epsilon}_{\rm GR;L}(k_{\sigma_l},\cdots,k_{\sigma_{k-1}},(\ell+P)^{h'_2},-\ell^{h'_1})\,{\cal A}^{\epsilon,\W\epsilon}_{\rm GR;R}(k_{\sigma_k},\cdots,k_{\sigma_{l-1}},-(\ell+P)^{\bar{h}'_2},\ell^{\bar{h}'_1})\,,
\eea
Using the relation
\bea
\sum_r\,(\epsilon_{P_+}^r)^\mu(\epsilon_{P_-}^r)^\nu\,{\cal A}_{\mu\nu}=\eta^{\mu\nu}\,{\cal A}_{\mu\nu}\,~~~~\label{ee-g}
\eea
for on-shell states, we see that the effect of applying ${\cal T}^\epsilon[\overline{+,-}]$ is turning both $(\epsilon_+\cdot\epsilon^r_{P_-})$ and $(\epsilon_-\cdot\epsilon^r_{P_+})$
to $1$ and removing the summation over $r$. In other words, the operator ${\cal T}^\epsilon[\overline{+,-}]$ turns $\sum_{h'_2}$
to $\sum_{h_2}$, and factorize into ${\cal T}^\epsilon[\overline{+,P_-}]$ and ${\cal T}^\epsilon[\overline{P_+,-}]$.
Obviously, the operator ${\cal T}^\epsilon[\overline{+,P_-}]$ acts on ${\cal A}^{\epsilon,\W\epsilon}_{\rm GR;R}$ and annihilates ${\cal A}^{\epsilon,\W\epsilon}_{\rm GR;L}$, while the operator ${\cal T}^\epsilon[\overline{P_+,-}]$ acts on  ${\cal A}^{\epsilon,\W\epsilon}_{\rm GR;L}$ and annihilates ${\cal A}^{\epsilon,\W\epsilon}_{\rm GR;R}$.
In the above argument, the fact
that each polarization vector appears once and only once in each term of the amplitude has been used. To continue, we observe
that ${\cal I}^\epsilon_{+\sigma_i\sigma_{i+1}}$ with $i\in\{k,\cdots,l-1\}$ acts on ${\cal A}^{\epsilon,\W\epsilon}_{\rm GR;R}$ and annihilates ${\cal A}^{\epsilon,\W\epsilon}_{\rm GR;L}$,
while those with $i\in\{l,\cdots,k-1\}$ acts on ${\cal A}^{\epsilon,\W\epsilon}_{\rm GR;L}$ and annihilates ${\cal A}^{\epsilon,\W\epsilon}_{\rm GR;R}$, based on whether the amplitude ${\cal A}^{\epsilon,\W\epsilon}_{\rm GR;L}$ or ${\cal A}^{\epsilon,\W\epsilon}_{\rm GR;R}$ includes the polarization vector $\epsilon_{\sigma_i}$.
Notice that one need to remove $k_-=-\ell$ in ${\cal A}^{\epsilon,\W\epsilon}_{\rm GR;L}$ and ${\cal A}^{\epsilon,\W\epsilon}_{\rm GR;R}$ by using the momentum conservation, to avoid the ambiguity that $\partial_{\epsilon_{\sigma_i}\cdot k_+}$ acts on $\epsilon_{\sigma_i}\cdot k_-$,
as what we did in section \ref{GR-YM-BAS}.
To regroup these insertion operators so that each operator can be interpreted appropriately, some treatments are in order.
For the operator ${\cal I}^\epsilon_{+\sigma_{l-1}\sigma_l}$, we rewrite it as
\bea
{\cal I}^\epsilon_{+\sigma_{l-1}\sigma_l}={\cal I}^\epsilon_{+\sigma_{l-1}P_-}+{\cal I}^\epsilon_{P_-\sigma_{l-1}\sigma_l}\,.
\eea
The operator ${\cal I}^\epsilon_{+\sigma_{l-1}P_-}$ can be interpreted as inserting the leg $\sigma_{l-1}$ between $+$ and $P_-$.
Since the momentum $k_{\sigma_l}$ enters ${\cal A}^{\epsilon,\W\epsilon}_{\rm GR;R}$ only through
$-P=\sum_{i=l}^{k-1}\,k_{\sigma_i}$, we have
\bea
\partial_{\epsilon_{\sigma_{l-1}}\cdot (-\ell-P)}\,{\cal A}^{\epsilon,\W\epsilon}_{\rm GR;R}(k_{\sigma_k},\cdots,k_{\sigma_{l-1}},-(\ell+P)^{\bar{h}'_2},\ell^{\bar{h}'_1})=\partial_{\epsilon_{\sigma_{l-1}}\cdot k_{\sigma_l}}\,{\cal A}^{\epsilon,\W\epsilon}_{\rm GR;R}(k_{\sigma_k},\cdots,k_{\sigma_{l-1}},-(\ell+P)^{\bar{h}'_2},\ell^{\bar{h}'_1})\,,
\eea
thus the operator ${\cal I}^\epsilon_{P_-\sigma_{l-1}\sigma_l}$ annihilates not only ${\cal A}^{\epsilon,\W\epsilon}_{\rm GR;L}$, but also ${\cal A}^{\epsilon,\W\epsilon}_{\rm GR;R}$.
Thus, when acting on ${\cal A}^{\epsilon,\W\epsilon}_{\rm GR;R}$, one can replace ${\cal I}^\epsilon_{+\sigma_{l-1}\sigma_l}$ by ${\cal I}^\epsilon_{+\sigma_{l-1}P_-}$.
Then, we arrive at the combinatory operator
\bea
\Big(\prod_{i=k}^{l-2}\,{\cal I}^\epsilon_{+\sigma_i\sigma_{i+1}}\Big)\,{\cal I}^\epsilon_{+\sigma_{l-1}P_-}\,{\cal T}^\epsilon[\overline{+,P_-}]
={\cal T}^\epsilon[\overline{+,\sigma_{k},\cdots,\sigma_{l-1},P_-}]\,,
\eea
which transmutes ${\cal A}^{\epsilon,\W\epsilon}_{\rm GR;R}$ to ${\cal A}^{\W\epsilon}_{{\rm YM};R}\big(\overline{(+)^{\bar{h}_1},\sigma_k,\cdots,\sigma_{l-1},(P_-)^{\bar{h}_2}}\big)$, and annihilates ${\cal A}^{\epsilon,\W\epsilon}_{\rm GR;L}$. Then we turn to the insertion operators ${\cal I}^\epsilon_{+\sigma_i\sigma_{i+1}}$ with $i\in\{l,\cdots,k-1\}$.
Since the momentum $k_-=-\ell$ has been removed via the momentum conservation, the momentum $\ell$ enters ${\cal A}^{\epsilon,\W\epsilon}_{\rm GR;L}$ only through $(\ell+P)$, thus, when applying $\partial_{\epsilon_{\sigma_i}\cdot \ell}$ to ${\cal A}^{\epsilon,\W\epsilon}_{\rm GR;L}$,
we have
\bea
\partial_{\epsilon_{\sigma_i}\cdot \ell}\,{\cal A}^{\epsilon,\W\epsilon}_{\rm GR;L}(k_{\sigma_l},\cdots,k_{\sigma_{k-1}},(\ell+P)^{h'_2},-\ell^{h'_1})=\partial_{\epsilon_{\sigma_i}\cdot (\ell+P)}\,{\cal A}^{\epsilon,\W\epsilon}_{\rm GR;L}(k_{\sigma_l},\cdots,k_{\sigma_{k-1}},(\ell+P)^{h'_2},-\ell^{h'_1})\,.
\eea
This equality allows us to interpret each ${\cal I}^\epsilon_{+\sigma_i\sigma_{i+1}}$ with $i\in\{l,\cdots,k-1\}$ as ${\cal I}^\epsilon_{P_+\sigma_i\sigma_{i+1}}$. But we still need to seek the correct interpretation of ${\cal I}^\epsilon_{P_+\sigma_{k-1}\sigma_k}$,
since $k_{\sigma_k}$ is included in ${\cal A}^{\epsilon,\W\epsilon}_{\rm GR;R}$  rather than in
${\cal A}^{\epsilon,\W\epsilon}_{\rm GR;L}$. Since $k_-=-\ell$ has been removed, we have
\bea
\partial_{\epsilon_{\sigma_i}\cdot k_{\sigma_k}}\,{\cal A}^{\epsilon,\W\epsilon}_{\rm GR;L}(k_{\sigma_l},\cdots,k_{\sigma_{k-1}},(\ell+P)^{h'_2},-\ell^{h'_1})=\partial_{\epsilon_{\sigma_i}\cdot k_-}\,{\cal A}^{\epsilon,\W\epsilon}_{\rm GR;L}(k_{\sigma_l},\cdots,k_{\sigma_{k-1}},(\ell+P)^{h'_2},-\ell^{h'_1})=0\,,
\eea
thus it is safe to replace ${\cal I}^\epsilon_{P_+\sigma_{k-1}\sigma_k}$ by ${\cal I}^\epsilon_{P_+\sigma_{k-1}-}$.
Consequently, we get another combinatory operator
\bea
\Big(\prod_{i=l}^{k-2}\,{\cal I}^\epsilon_{P_+\sigma_i\sigma_{i+1}}\Big)\,{\cal I}^\epsilon_{P_+\sigma_{k-1}-}\,{\cal T}^\epsilon[\overline{P_+,-}]
={\cal T}^\epsilon[\overline{P_+,\sigma_{l},\cdots,\sigma_{k-1},-}]\,,
\eea
which transmutes ${\cal A}^{\epsilon,\W\epsilon}_{\rm GR;L}$ to ${\cal A}^{\W\epsilon}_{{\rm YM};L}\big(\overline{(P_+)^{h_2},\sigma_l,\cdots,\sigma_{k-1},(-)^{h_1}}\big)$, and annihilates ${\cal A}^{\epsilon,\W\epsilon}_{\rm GR;R}$. Based on the above discussion, we conclude that under the on-shell condition $(l+P)^2=0$,
the operator ${\cal T}^\epsilon[\overline{+,\sigma_k,\cdots,\sigma_{k-1},-}]$ factorizes into
\bea
{\cal O}_L={\cal T}^\epsilon[\overline{P_+,\sigma_{l},\cdots,\sigma_{k-1},-}]\,,~~~~~~~~{\cal O}_R={\cal T}^\epsilon[\overline{+,\sigma_{k},\cdots,\sigma_{l-1},P_-}]\,,
\eea
and transmutes the factorization of GR amplitude to the factorization of color ordered YM amplitude.

\subsection{Operator ${\cal L}^\epsilon_\circ\,{\cal D}$}
\label{OP-2}

The next example is the factorization of operator ${\cal L}^\epsilon_\circ\,{\cal D}$. We will demonstrate it by considering the relation
${\bf I}^{\W\epsilon}_{\rm BI}={\cal L}^\epsilon_\circ\,{\cal D}\,{\bf I}^{\epsilon,\W\epsilon}_{\rm GR}$. In the case under consideration,
\eref{rewitten-reverse-forward} becomes
\bea
\lim_{\ell^2\to 0}\,\ell^2\,{\cal L}^\epsilon_{\circ C}\,{\cal D}\,{\bf I}^{\epsilon,\W\epsilon}_{\rm GR}({\pmb H}_n)&=&\sum_{h_1}\,{\cal L}^\epsilon\,{\cal T}^\epsilon[\overline{+,-}]\,{\cal A}^{\epsilon,\W\epsilon}_{\rm GR}({\pmb H}_n+\{-\ell^{h'_1},\ell^{\bar{h}'_1}\})\nn
&=&\sum_{h_1}\,{\cal A}^{\W\epsilon}_{\rm BI}({\pmb P}_n+\{-\ell^{h_1},\ell^{\bar{h}_1}\})\,.
\eea
Doing the second step in \eref{2-steps}, then we need to show
\bea
& &\lim_{(\ell+P)^2\to 0}\,(\ell+P)^2\,{\cal L}^\epsilon\,{\cal T}^\epsilon[\overline{+,-}]\,{\cal A}^{\epsilon,\W\epsilon}_{\rm GR}({\pmb H}_n+\{-\ell^{h'_1},\ell^{\bar{h}'_1}\})\nn
&=&\sum_{h_2}\,\Big({\cal O}_L\,{\cal A}^{\epsilon,\W\epsilon}_{{\rm GR};L}(k_{\sigma_l},\cdots,k_{\sigma_{k-1}},(\ell+P)^{h'_2},-\ell^{h'_1})\Big)\,\Big({\cal O}_R\,{\cal A}^{\epsilon,\W\epsilon}_{{\rm GR};R}(k_{\sigma_k},\cdots,k_{\sigma_{l-1}},-(\ell+P)^{\bar{h}'_2},\ell^{\bar{h}'_1})\Big)\,,~~~~\label{fac-LT}
\eea
which is just the tree
level factorization
\bea
& &\lim_{(\ell+P)^2\to 0}\,(\ell+P)^2\,{\cal A}^{\W\epsilon}_{\rm BI}({\pmb P}_n+\{-\ell^{h_1},\ell^{\bar{h}_1}\})\nn
&=&\sum_{h_2}\,{\cal A}^{\W\epsilon}_{{\rm BI};L}(k_{\sigma_l},\cdots,k_{\sigma_{k-1}},(\ell+P)^{h_2},-\ell^{h_1})\,{\cal A}^{\W\epsilon}_{{\rm BI};R}(k_{\sigma_k},\cdots,k_{\sigma_{l-1}},-(\ell+P)^{\bar{h}_2},\ell^{\bar{h}_1})\,.
\eea
Again, the factorization of the operator ${\cal L}^\epsilon\,{\cal T}^\epsilon[\overline{+,-}]$ in \eref{fac-LT} is ensured by the factorization of tree amplitude.

It is straightforward to understand the factorization of the operator ${\cal L}^\epsilon\,{\cal T}^\epsilon[\overline{+,-}]$ at tree level. In the previous subsection, we proved that the operator ${\cal T}^\epsilon[\overline{+,-}]$ factorizes into ${\cal T}^\epsilon[\overline{+,P_-}]$ and ${\cal T}^\epsilon[\overline{P_+,-}]$
under the on-shell condition $(\ell+P)^2=0$. One can regroup ${\cal L}^\epsilon$ to be ${\cal L}^\epsilon={\cal L}^\epsilon_L\,{\cal L}^\epsilon_R$
directly,
by considering whether ${\cal A}^{\epsilon,\W\epsilon}_{\rm GR;L}$ or ${\cal A}^{\epsilon,\W\epsilon}_{\rm GR;R}$ contains the polarization vector $\epsilon_{\sigma_i}$. We emphasize that two definitions
${\cal L}^\epsilon$ and $\bar{{\cal L}}^\epsilon$ in \eref{defin-L} leads to the same factorization. Notice that when considering $\bar{{\cal L}}^\epsilon$,
for the operator $(k_i\cdot k_j)\partial_{\epsilon_i\cdot \epsilon_j}$ with $\epsilon_i\in{\cal A}^{\epsilon,\W\epsilon}_{{\rm GR};L}$ and $\epsilon_j\in{\cal A}^{\epsilon,\W\epsilon}_{{\rm GR};R}$,
one can not use the technic of separating ${\cal T}^\epsilon[\overline{+,-}]$ into ${\cal T}^\epsilon[\overline{P_+,-}]$ and ${\cal T}^\epsilon[\overline{+,P_-}]$
to separate $\partial_{\epsilon_i\cdot \epsilon_j}$. The reason is, the summation $\sum_r\,(\epsilon_{P_+}^r)^\mu(\epsilon_{P_-}^r)^\nu$
has been removed when separating ${\cal T}^\epsilon[\overline{+,-}]$. Thus, the operator $(k_i\cdot k_j)\partial_{\epsilon_i\cdot \epsilon_j}$
across ${\cal A}^{\epsilon,\W\epsilon}_{{\rm GR};L}$ and ${\cal A}^{\epsilon,\W\epsilon}_{{\rm GR};R}$ annihilates both two amplitudes.
After regrouping ${\cal L}^\epsilon$, we find that the operator
${\cal L}^\epsilon\,{\cal T}^\epsilon[\overline{+,-}]$ factorizes into
\bea
{\cal O}_L={\cal L}^\epsilon_L\,{\cal T}^\epsilon[\overline{P_+,-}]\,,~~~~~~~~{\cal O}_R={\cal L}^\epsilon_R\,{\cal T}^\epsilon[\overline{+,P_-}]\,.
\eea
%

\subsection{Operator ${\cal T}^\epsilon_{X_{2m}}\,({\cal D}+1)$}
\label{OP-3}

The final example is the operator ${\cal T}^\epsilon_{X_{2m}}\,({\cal D}+1)$, which links the $1$-loop GR and EM Feynman integrands together
as ${\bf I}^{\epsilon,\W\epsilon}_{\rm EM}({\pmb P}_{2m};{\pmb H}_{n-2m})={\cal T}^\epsilon_{X_{2m}}\,({\cal D}+1)\,{\bf I}^{\epsilon,\W\epsilon}_{\rm GR}({\pmb H}_{n})$. The factorization of the operator ${\cal T}^\epsilon_{{\cal X}_{2m}}\,(N\,{\cal D}+1)$ can be discussed similarly.
For latter convenience, let us assume that the external gravitons in the set $\{1,\cdots,2m\}$ are turned to photons by the operator
${\cal T}^\epsilon_{X_{2m}}\,({\cal D}+1)$.
For the current example,
the new situation occurs. A special cut should be recognized not only by the momenta channel, but also what kind of cut virtual particle.
We need to discuss the following three cases, which are cutting two gravitons, cutting one graviton and one photon, as well as cutting two photons.

As before, we start by doing the first step in \eref{2-steps}. Using \eref{I-EM} and \eref{OP-EM}, we know that
\bea
\lim_{\ell^2\to 0}\,\ell^2\,{\cal T}^\epsilon_{X_{2m}}\,({\cal D}+1)\,{\bf I}^{\epsilon,\W\epsilon}_{\rm GR}({\pmb H}_{n})&=&\sum_{h'_1}\,{\cal T}^\epsilon_{X_{2m}}\,{\cal A}^{\epsilon,\W\epsilon}_{\rm GR}({\pmb H}_n+\{-\ell^{h'_1},\ell^{\bar{h}'_1}\})\nn
&=&\sum_{h'_1}\,{\cal A}^{\epsilon,\W\epsilon}_{\rm EM}({\pmb P}_{2m};{\pmb H}_{n-2m}+\{-\ell^{h'_1},\ell^{\bar{h}'_1}\})\,,~~~~\label{case-1}
\eea
if we cut two gravitons. We also have
\bea
\lim_{\ell^2\to 0}\,\ell^2\,{\cal T}^\epsilon_{X_{2m}}\,({\cal D}+1)\,{\bf I}^{\epsilon,\W\epsilon}_{\rm GR}({\pmb H}_{n})&=&\sum_{h_1}\,{\cal T}^\epsilon_{X_{2m+2}}\,{\cal A}^{\epsilon,\W\epsilon}_{\rm GR}({\pmb H}_n+\{-\ell^{h'_1},\ell^{\bar{h}'_1}\})\nn
&=&\sum_{h_1}\,{\cal A}^{\epsilon,\W\epsilon}_{\rm EM}({\pmb P}_{2m}+\{-\ell^{h_1},\ell^{\bar{h}_1}\};{\pmb H}_{n-2m})\,,~~~~\label{case-2}
\eea
if we cut two photons.

Now we do the second step in \eref{2-steps}. For the first case \eref{case-1}, if we cut two gravitons in the second step, all
${\cal T}^\epsilon[\overline{a,b}]$ with $\epsilon_a\in{\cal A}^{\epsilon,\W\epsilon}_{{\rm GR};L}$ and $\epsilon_b\in{\cal A}^{\epsilon,\W\epsilon}_{{\rm GR};R}$ annihilates both two amplitudes, thus the effective
operator becomes
\bea
{\cal T}^\epsilon_{X_{2m}}|_{\rm effective}&=&\Big(\sum_{\rho_L}\,\prod_{i_k,j_k\in\rho_L}\,{\cal T}^\epsilon[\overline{i_k,j_k}]\Big)\,
\Big(\sum_{\rho_R}\,\prod_{i_l,j_l\in\rho_R}\,{\cal T}^\epsilon[\overline{i_l,j_l}]\Big)\nn
&=&{\cal T}^\epsilon_{X_{m_L}}\,{\cal T}^\epsilon_{X_{m_R}}\,,
\eea
where $m_L$ and $m_R$ denote the number of photons in ${\cal A}_{{\rm EM};L}$ and ${\cal A}_{{\rm EM};R}$ respectively, and satisfy $m_L+m_R=2m$. More explicitly, we have assumed that the operator ${\cal T}^\epsilon_{X_{2m}}$
turns external gravitons belong to $\{1,\cdots,2m\}$ to photons, and now we divide $\{1,\cdots,2m\}$ into $\{a_1,\cdots,a_{m_L}\}$ and $\{b_1,\cdots,b_{m_R}\}$, where the external graviton labeled by $a_i$ is included in ${\cal A}^{\epsilon,\W\epsilon}_{{\rm GR};L}$, while that labeled by $b_i$ is included in ${\cal A}^{\epsilon,\W\epsilon}_{{\rm GR};R}$. Partitions $\rho_L$ and $\rho_R$ are understood as partitions for $\{a_1,\cdots,a_{m_L}\}$ and $\{b_1,\cdots,b_{m_R}\}$, respectively. Thus, for this case, the operator ${\cal T}^\epsilon_{X_{2m}}$ factorizes
into
\bea
{\cal O}_L={\cal T}^\epsilon_{X_{m_L}}\,,~~~~~~~~{\cal O}_R={\cal T}^\epsilon_{X_{m_R}}\,.~~~~\label{fac-1}
\eea

If we cut two photons in the second step, the effective part of the operator ${\cal T}^\epsilon_{X_{2m}}$ should contain one and only one ${\cal T}^\epsilon[\overline{a,b}]$ with $\epsilon_a\in{\cal A}^{\epsilon,\W\epsilon}_{{\rm GR};L}$ and $\epsilon_b\in{\cal A}^{\epsilon,\W\epsilon}_{{\rm GR};R}$, and we can use \eref{ee-g} to see that ${\cal T}^\epsilon[\overline{a,b}]$
factorizes into ${\cal T}^\epsilon[\overline{a,P_+}]$ and ${\cal T}^\epsilon[\overline{P_-,b}]$. Notice that such decomposition can be used only once, as discussed in the previous subsection. Then we find
\bea
{\cal T}^\epsilon_{X_{2m}}|_{\rm effective}&=&\sum_{\substack{a\in\{a_1,\cdots,a_{2m_L}\}\\b\in\{b_1,\cdots,b_{2m_R}\}}}\,\Big({\cal T}^\epsilon[\overline{a,P_+}]\,\sum_{\rho^a_L}\,\prod_{i_k,j_k\in\rho^a_L}\,{\cal T}^\epsilon[\overline{i_k,j_k}]\Big)\,
\Big({\cal T}^\epsilon[\overline{P_-,b}]\,\sum_{\rho^b_R}\,\prod_{i_l,j_l\in\rho^b_R}\,{\cal T}^\epsilon[\overline{i_l,j_l}]\Big)\nn
&=&{\cal T}^\epsilon_{X_{m_L+1}}\,{\cal T}^\epsilon_{X_{m_R+1}}\,.
\eea
Here $\rho^a_L$ and $\rho^b_R$ are understood as partitions for $\{a_1,\cdots,a_{m_L}\}\setminus a$ and $\{b_1,\cdots,b_{m_R}\}\setminus b$, respectively. Thus the operator ${\cal T}^\epsilon_{X_{2m}}$ factorizes into
\bea
{\cal O}_L={\cal T}^\epsilon_{X_{m_L+1}}\,,~~~~~~~~{\cal O}_R={\cal T}^\epsilon_{X_{m_R+1}}\,.~~~~\label{fac-2}
\eea

Similar discussion holds for the second case \eref{case-2}. For the second case, if we cut two gravitons at the second step,
we find
\bea
{\cal T}^\epsilon_{X_{2m+2}}|_{\rm effective}&=&\Big(\sum_{\rho'_L}\,\prod_{i_k,j_k\in\rho'_L}\,{\cal T}^\epsilon[\overline{i_k,j_k}]\Big)\,
\Big(\sum_{\rho'_R}\,\prod_{i_l,j_l\in\rho'_R}\,{\cal T}^\epsilon[\overline{i_l,j_l}]\Big)\nn
&=&{\cal T}^\epsilon_{X'_{m_L+1}}\,{\cal T}^\epsilon_{X'_{m_R+1}}\,.~~~~\label{fac-3}
\eea
Here we use the notations ${\cal T}^\epsilon_{X'_{m_L+1}}$ and ${\cal T}^\epsilon_{X'_{m_R+1}}$ to distinguish them
from ${\cal T}^\epsilon_{X_{m_L+1}}$ and ${\cal T}^\epsilon_{X_{m_R+1}}$ in \eref{fac-2}.
This formula should be understood as follows. The operator ${\cal T}^\epsilon_{X_{2m+2}}$ turns external gravitons in
$\{1,\cdots,2m,+,-\}$ to photons, and we divide $\{1,\cdots,2m\}$ into $\{a_1,\cdots,a_{m_L}\}$ and $\{b_1,\cdots,b_{m_R}\}$
as before. Then $\rho'_L$ and $\rho'_R$ are partitions for $\{a_1,\cdots,a_{m_L},-\}$ and $\{b_1,\cdots,b_{m_R},+\}$, respectively.
The factorized formula in \eref{fac-3} is the similar as that in \eref{fac-2}. This similarity is quite natural, since from the unitarity cut point of view, both two cases are cutting one graviton and one photon. Cutting two photons at the second step gives
\bea
{\cal T}^\epsilon_{X_{2m+2}}|_{\rm effective}&=&\sum_{\substack{a\in\{a_1,\cdots,a_{m_L},-\}\\b\in\{b_1,\cdots,b_{m_R},+\}}}\,\Big({\cal T}^\epsilon[\overline{a,P_+}]\,\sum_{\rho'^a_L}\,\prod_{i_k,j_k\in\rho'^a_L}\,{\cal T}^\epsilon[\overline{i_k,j_k}]\Big)\,
\Big({\cal T}^\epsilon[\overline{P_-,b}]\,\sum_{\rho'^b_R}\,\prod_{i_l,j_l\in\rho'^b_R}\,{\cal T}^\epsilon[\overline{i_l,j_l}]\Big)\nn
&=&{\cal T}^\epsilon_{X_{m_L+2}}\,{\cal T}^\epsilon_{X_{m_R+2}}\,,
\eea
thus the operator ${\cal T}^\epsilon_{X_{2m+2}}$ factorizes into ${\cal T}^\epsilon_{X_{m_L+2}}$ and ${\cal T}^\epsilon_{X_{m_R+2}}$.
Here $\rho'^a_L$ and $\rho'^b_R$ are partitions of $\{a_1,\cdots,a_{m_L},-\}\setminus a$ and $\{b_1,\cdots,b_{m_R},+\}\setminus b$, respectively.

\section{Summary and discussions}
\label{secconclu}

%
\begin{table}[!h]
\begin{center}
\begin{tabular}{c|c|c}
Feynman integrand& ${\cal O}^\epsilon_\circ$  & ${\cal O}^{\W\epsilon}_\circ$ \\
\hline
${\bf I}_{{\rm GR}}^{\epsilon,\W\epsilon}(\pmb{H}_n)$ & $\mathbb{I}$ & $\mathbb{I}$  \\
${\bf I}_{{\rm ssEYM}}^{\epsilon,\W\epsilon}(\overline{\pmb{\sigma}}_m;\pmb{H}_{n-m})$ & $\mathbb{I}$ & ${\cal T}^{\W\epsilon}_\circ[\overline{\pmb{\sigma}}_m]$  \\
${\bf I}_{{\rm EMf}}^{\epsilon,\W\epsilon}(\pmb{P}_{2m};\pmb{H}_{n-2m})$ & $\mathbb{I}$ & ${\cal T}^{\W\epsilon}_{{\cal X}_{2m}}(N\,\W{\cal D}+1)$  \\
${\bf I}_{{\rm EM}}^{\epsilon,\W\epsilon}(\pmb{P}_{2m};\pmb{H}_{n-2m})$ & $\mathbb{I}$ & ${\cal T}^{\W\epsilon}_{X_{2m}}(\W{\cal D}+1)$  \\
${\bf I}_{{\rm BI}}^\epsilon(\pmb{P}_n)$ & $\mathbb{I}$ & ${\cal L}^{\W\epsilon}_\circ\,\W{\cal D}$ \\
${\bf I}_{{\rm YM}}^\epsilon(\overline{\pmb{\sigma}}_n)$ & $\mathbb{I}$ & ${\cal T}^{\W\epsilon}_\circ[\overline{\pmb{\sigma}}_n]$  \\
${\bf I}_{{\rm ssYMS}}^{\W\epsilon}(\overline{\pmb{\sigma}}_m;\pmb{G}_{n-m}|\overline{\pmb{\sigma}'}_n)$ & ${\cal T}^{\epsilon}_\circ[\overline{\pmb{\sigma}'}_n]$ & ${\cal T}^{\W\epsilon}_\circ[\overline{\pmb{\sigma}}_m]$ \\
${\bf I}_{{\rm NLSM}}(\overline{\pmb{\sigma}'}_n)$ & ${\cal T}^{\epsilon}_\circ[\overline{\pmb{\sigma}'}_n]$ & ${\cal L}^{\W\epsilon}_\circ\,\W{\cal D}$ \\
${\bf I}_{{\rm BAS}}(\overline{\pmb{\sigma}}_n|\overline{\pmb{\sigma}'}_n)$ &  ${\cal T}^{\epsilon}_\circ[\overline{\pmb{\sigma}'}_n]$ & ${\cal T}^{\W\epsilon}_\circ[\overline{\pmb{\sigma}}_n]$ \\
${\bf I}_{{\rm DBI}}^{\W\epsilon}(\pmb{S}_{2m};\pmb{P}_{n-2m})$ &${\cal L}^{\epsilon}_\circ\,{\cal D}$ & ${\cal T}^{\W\epsilon}_{{\cal X}_{2m}}(N\,\W{\cal D}+1)$ \\
${\bf I}_{{\rm SG}}(\pmb{S}_n)$ &  ${\cal L}^{\epsilon}_\circ\,{\cal D}$ & ${\cal L}^{\W\epsilon}_\circ\,\W{\cal D}$ \\
\end{tabular}
\end{center}
\caption{\label{tab:unifying-loop}Unifying relations at $1$-loop level.}
\end{table}

\begin{figure}
  \centering
  \includegraphics[width=12cm]{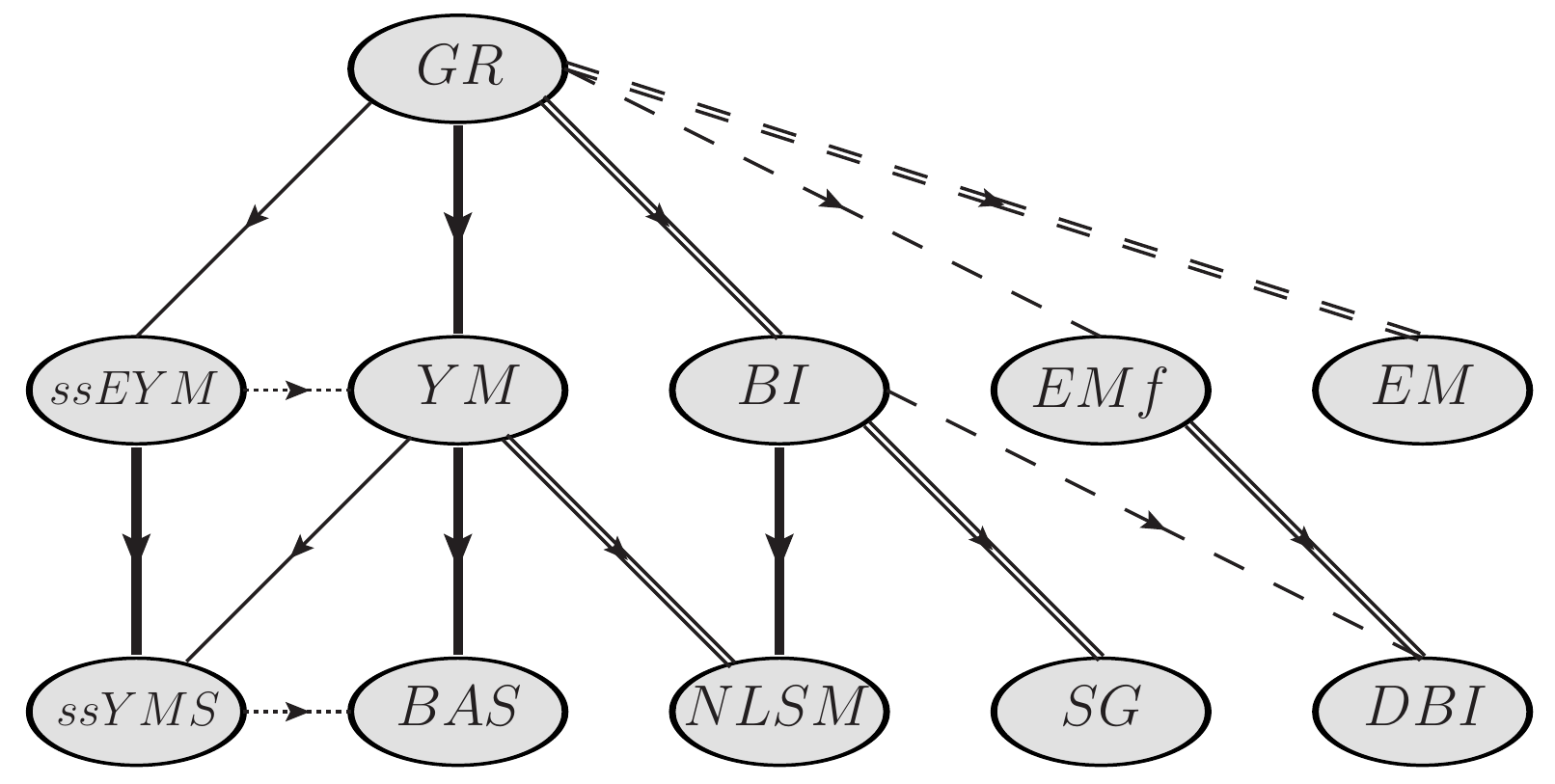} \\
  \caption{Unified web for $1$-loop Feynman integrands. The bold straight line represents the operator ${\cal T}_\circ[\overline{\pmb{\sigma}}_n]$, the straight line represents the operator ${\cal T}_\circ[\overline{\pmb{\sigma}}_m]$, the double straight line represents the operator ${\cal L}\,{\cal D}$, the dashed line represents the operator ${\cal T}_{{\cal X}_{2m}}(N\,{\cal D}+1)$, the double dashed line represents the operator ${\cal T}_{X_{2m}}({\cal D}+1)$, the thin dashed line represents the insertion operators.}\label{unifyingweb}
\end{figure}

In this paper, we have constructed the $1$-loop level differential operators which transmute the $1$-loop GR Feynman integrand to the Feynman integrands of various theories, include Einstein-Yang-Mills theory, Einstein-Maxwell theory, pure Yang-Mills theory, Yang-Mills-scalar theory, Born-Infeld theory,
Dirac-Born-Infeld theory, bi-adjoint scalar theory, non-linear sigma model,
as well as special Galileon theory. Similar as the tree level formula \eref{fund-uni-diff}, the $1$-loop level relations can be summarized as
\bea
{\bf I}={\cal O}^\epsilon_{\circ}\,{\cal O}^{\W\epsilon}_{\circ}\,{\bf I}^{\epsilon,\W\epsilon}_{\rm GR}\,.
\eea
Operators ${\cal O}^\epsilon_{\circ}$ and ${\cal O}^{\W\epsilon}_{\circ}$ for different theories are listed in Table.\ref{tab:unifying-loop}.
Comparing Table.\ref{tab:unifying-loop} with Table.\ref{tab:unifying}, we see that the $1$-loop level relations are paralleled to the tree level
relations.
Based on the relations in Table.\ref{tab:unifying-loop}, one can construct the unified web for these theories, as shown in Fig.\ref{unifyingweb}.
In this web, different kinds of lines represents different operators. Along a line from theory $A$ to theory $B$, one can generate the $1$-loop Feynman integrand ${\bf I}_B$, by applying the corresponding operator to ${\bf I}_A$. For example, there is a double straight line from EMf to DBI,
thus we have the relation
\bea
{\bf I}^{\W\epsilon}_{\rm DBI}({\pmb S}_{2m};{\pmb P}_{n-2m})={\cal L}^\epsilon_\circ\,{\cal D}\,{\bf I}^{\epsilon,\W\epsilon}_{\rm EMf}({\pmb P}_{2m};{\pmb H}_{n-2m})\,.
\eea

We also discussed the factorization of differential operators under the well known unitarity cut. Our result shows that the $1$-loop level operator factorizes into two tree level pieces, paralleled to the factorization of amplitudes. As a by product, we also found the mechanism of the factorization of tree level operators. These properties propose an alternative way to verify the operators: applying the tree level operator to the lowest points tree level amplitudes. A natural conjecture is, maybe one can construct a general operator recursively, from the lowest points operator, similar as the well known on-shell recursion relation for constructing amplitudes.

For the EYM theory, in this paper we only consider the single trace Feynman integrands with a virtual gluon running in the loop, and similar does the YMS theory. We have not discussed the more general case, due to some technical difficulties those we do not know how to overcome. The general multiple traces case with arbitrary virtual particle in the loop is left to be the future direction.

The basic idea used in this paper is to find the operator ${\cal O}_\circ$ satisfying ${\cal O}_\circ\,{\cal F}\,\bullet={\cal F}\,{\cal O}\,\bullet$, where ${\cal O}$ is the already known tree level operator. This idea based on the validity of the forward limit method. However, as mentioned in section \ref{secintro}, the forward limit method is not the only candidate for generating the $1$-loop CHY formula. Thus, how to reveal relations in Table.\ref{tab:unifying-loop} and Fig.\ref{unifyingweb} in the ambitwistor string framework is an interesting question.
The Feynman integrands provided by the current $1$-loop CHY formula in the sense of ${\cal Q}$-cut \cite{Baadsgaard:2015twa,Huang:2015cwh}. They include propagators linear in $\ell$ rather than quadratic, and are related to the standard ones via the partial fraction identity. Until now we do not have a systematical way to construct the
CHY formula which leads to Feynman integrands with propagators quadratic in $\ell$. Searching the differential operators link Feynman integrands with quadratic propagators together, via a path which is independent of the CHY formula, is another interesting challenge.

\section*{Acknowledgments}

The author would thank Prof. Bo Feng, Xiaodi-Li, Chang Hu, Yaobo Zhang, Tingfei Li, especially Prof. Bo Feng and Xiaod- Li, for helpful discussions and valuable suggestions.
This
work is supported by Chinese NSF funding under
contracts No.11805163, as well as NSF of Jiangsu Province under Grant No.BK20180897.


\end{document}